\documentclass[onecolumn,11pt]{article}
\usepackage[top=1in, bottom=1in, left=1in, right=1in]{geometry}
\usepackage{amsmath,amsfonts,amscd,amssymb}
\usepackage{graphicx}
\usepackage{epstopdf}
\usepackage{overpic}
\usepackage{cancel}
\usepackage{rotating}
\usepackage{url}
\usepackage{caption}
\usepackage{color}
\usepackage{rotating}
\usepackage{multirow}
\usepackage{wrapfig}
\usepackage{mathtools}
\usepackage{subeqnarray}
\usepackage{setspace}
\usepackage{palatino} 
\usepackage[bottom,flushmargin,hang,multiple]{footmisc}
\usepackage{lipsum}
\newcommand\blfootnote[1]{%
  \begingroup
  \renewcommand\thefootnote{}\footnote{#1}%
  \addtocounter{footnote}{-1}%
  \endgroup
}

\usepackage[numbers,sort&compress]{natbib}

\definecolor{offlinered}{RGB}{115, 0, 0}
\definecolor{onlineblue}{RGB}{4, 79, 149}

\newcommand{\R}{\rm I\!R}
\renewcommand{\Re}{\textrm{Re}}
\newcommand{\secref}[1]{\S\ref{#1}}

\graphicspath{{figs_v2/}}

\title{\LARGE{\vspace{-.5in}\textbf{Robust flow reconstruction from limited measurements\\ via sparse representation}}\vspace{-.15in}}
\author{\normalsize{Jared L. Callaham$^{1*}$, Kazuki Maeda$^2$, and Steven L. Brunton$^2$}\\
\footnotesize{$^1$ Department of Applied Mathematics, University of Washington, Seattle, WA 98195, United States}\\
\footnotesize{$^2$ Department of Mechanical Engineering, University of Washington, Seattle, WA 98195, United States\vspace{-.1in}}
}
\date{}
\begin{document}
\maketitle

\blfootnote{$^*$ Corresponding author (jc244@uw.edu).\\ \noindent \textbf{Matlab code:}  github.com/jcallaham/robust-flow-reconstruction}

\vspace{-.2in}
\begin{abstract}
In many applications it is important to estimate a fluid flow field from limited and possibly corrupt measurements. Current methods in flow estimation often use least squares regression to reconstruct the flow field, finding the minimum-energy solution that is consistent with the measured data. However, this approach may be prone to overfitting and sensitive to noise. To address these challenges we instead seek a sparse representation of the data in a library of examples.  
Sparse representation has been widely used for image recognition and reconstruction, and it is well-suited to structured data with limited, corrupt measurements.
We explore sparse representation for flow reconstruction on a variety of fluid data sets with a wide range of complexity, including vortex shedding past a cylinder at low Reynolds number, a mixing layer, and two geophysical flows.  In addition, we compare several measurement strategies and consider various types of noise and corruption over a range of intensities.  We find that sparse representation has considerably improved estimation accuracy and robustness to noise and corruption compared with least squares methods.  We also introduce a sparse estimation procedure on local spatial patches for complex multiscale flows that preclude a global sparse representation.  Based on these results, sparse representation is a promising framework for extracting useful information from complex flow fields with realistic measurements.  \\

\end{abstract}

\section{Introduction}\label{sec:introduction}
Estimating the structure of a flow field from limited and noisy measurements is an important challenge in many engineering applications.  For example, accurate estimation is central to active flow control \citep{Noack2011book,King2014book,Brunton2015,Sipp2016amr}, which has the potential to advance next-generation technology, ranging from fuel-efficient, low-drag automobiles~\citep{Pfeiffer2016} to high-efficiency turbines~\citep{Strom2017} and internal combustion engines~\citep{Maurya2017}. The ability to reconstruct important flow features from restricted observations is also critical in applications as diverse as cardiac bloodflow modeling~\citep{Yakhot2008, Sankaran2012}, ship wake identification~\citep{Graziano2016}, and climate science~\citep{Kalnay2003}.  All of these applications rely on estimating the structure of complex fluid flows based on limited measurements. 
This work focuses on addressing this challenge using modern techniques from machine learning and sparse representation~\citep{Wright2009}, which have recently been applied to flow field classification~\citep{Bright2013,Brunton2014,Bright2016,Kramer2017}.

Modern experimental methods and the increasing scale and resolution of numerical simulations have led to an abundance of fluid flow data. Although we are able to achieve unprecedented fidelity in measurement and simulation in lab settings, in applications we are typically limited to a few noisy sensors.  The challenge in flow field estimation is thus synthesizing the profusion of offline data and limited, unreliable online information.
This synthesis relies on learning and representing the essential structure of the flow field by leveraging the physical behavior observed in past data.    
Fluid mechanics is not unique in having a wealth of data, however, and recent years have seen the rapid development of revolutionary machine learning techniques to leverage big data, particularly in image processing~\citep{Krizhevsky2012nips,Lecun2015nature}.  
Since flow field data is often discretized on a grid, many of these techniques can be applied to fluid mechanics with only minor modifications, for example for flow field classification~\citep{Bright2013} and estimation~\citep{Yu2018}.


A common model-free approach to flow field reconstruction is to represent the field as a linear combination of modes in a library, such as empirical eigenfunctions from proper orthogonal decomposition (POD)~\citep{Sirovich1987, Berkooz1993} or dynamic mode decomposition (DMD)~\citep{Schmid2010, Rowley2009, Tu2014a, Kutz2016book}.  
Gappy POD was introduced by Everson and Sirovich~\citep{Everson1995} to repair corrupted or missing data, and was adapted to flow field reconstruction by Bui-Thanh et al.~\citep{Bui-Thanh2004}.  
This method has been used to reconstruct unsteady flow fields around a cylinder~\citep{Venturi2004} and an airfoil~\citep{Wilcox2006}, arterial blood flow data~\citep{Yakhot2008}, and low-dimensional ocean velocity and temperature fields~\citep{Yildrim2009}. Podvin et al. \citep{Podvin2005} used a similar method to estimate POD coefficients for 3D cavity flow  from 2D particle image velocimetry (PIV) data.
Other leading methods for flow field estimation include stochastic estimation and model-based observers, which are discussed in more detail in section~\ref{sec:background}. 
However, these majority of these approaches are based on least-squares regression, which may be prone to overfitting and  sensitive to noise.
Furthermore, although these methods minimize the kinetic energy deviation between the predicted and actual measurements, this does not guarantee that the reconstruction will be globally optimal.

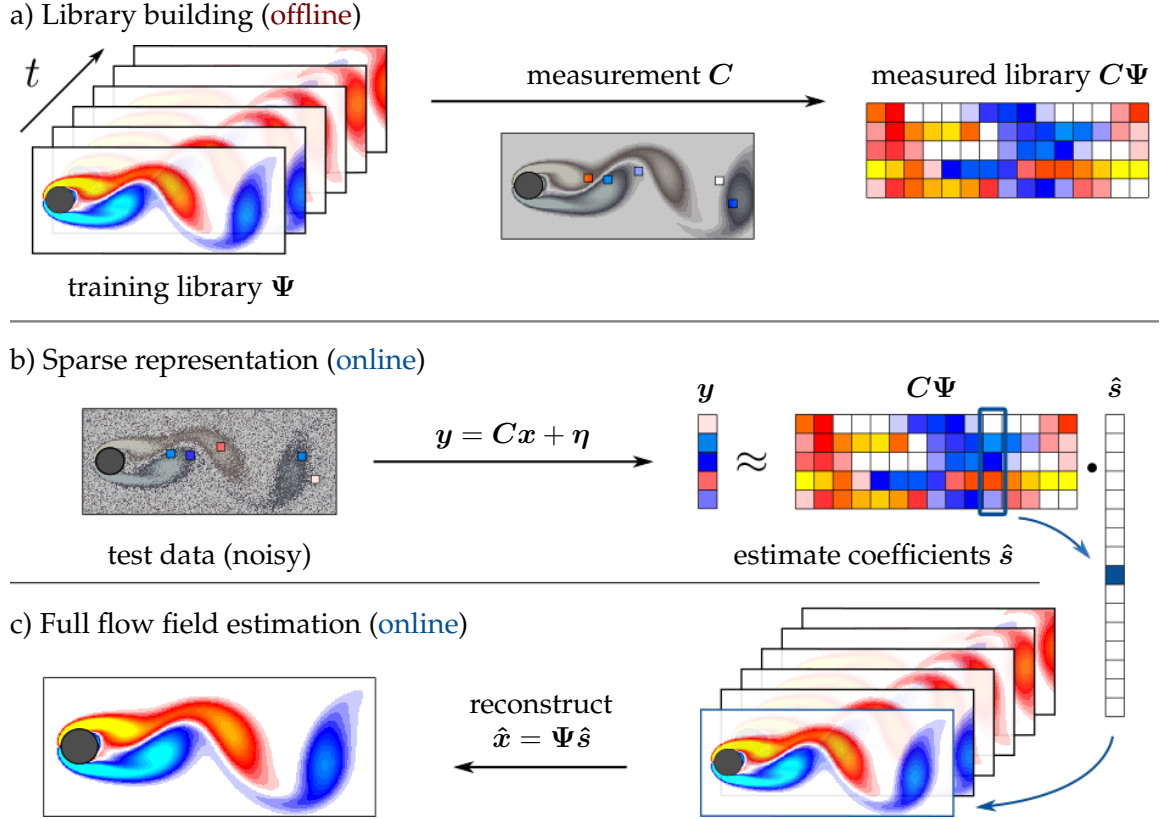
\begin{figure}
	\centering
	\vspace{.3in}
	\begin{overpic}[width=0.925\linewidth]{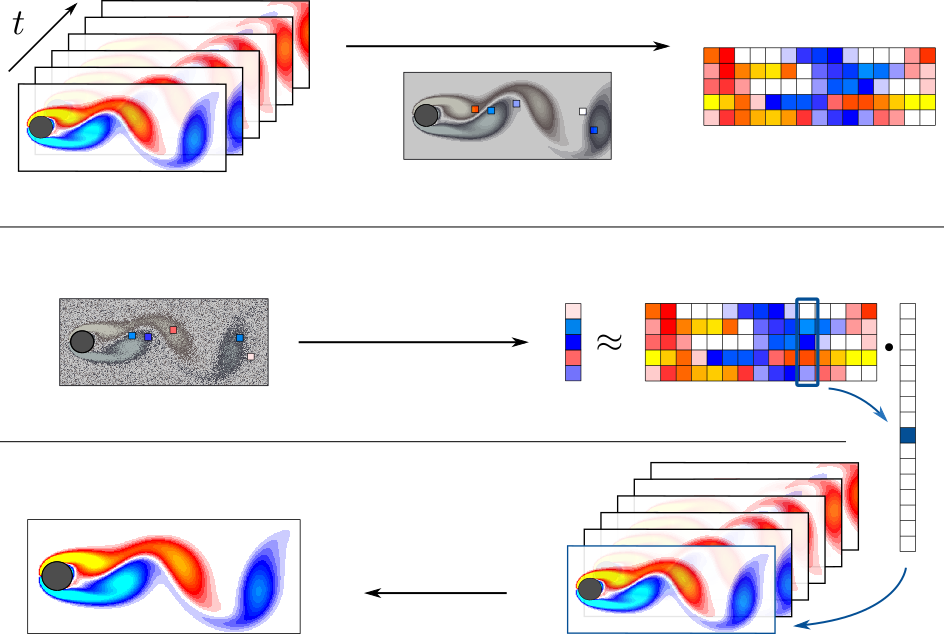}
		\put(0, 69){a) Library building (\textcolor{offlinered}{offline})}
		\put(0, 39){b) Sparse representation (\textcolor{onlineblue}{online})}
		\put(0, 16){c) Full flow field estimation (\textcolor{onlineblue}{online})}
		\put(5, 45.3){training library $\boldsymbol{\Psi}$}
		\put(45, 63.8){measurement $ \boldsymbol{C} $}
		\put(75, 63.8){measured library $ \boldsymbol{C}\boldsymbol \Psi $}
		\put(8.5, 22){test data (noisy)}
		\put(37, 32.5){$ \boldsymbol{ y}  = \boldsymbol{C}\boldsymbol{x} + \boldsymbol{\eta} $}
		\put(59.8, 36.5){$ \boldsymbol y $}
		\put(78, 36.5){$ \boldsymbol{C} \boldsymbol \Psi $}
		\put(95.5, 36.5){$ \boldsymbol {\hat{s}} $}
		\put(63, 22){estimate coefficients $ \boldsymbol {\hat{s}} $}
		\put(40, 9){reconstruct}
		\put(42, 6){$ \boldsymbol {\hat{x}} = \boldsymbol{\Psi \hat{s}} $}
	\end{overpic}
	\vspace{.1in}
	\caption{Flow field reconstruction process using sparse representation. In offline library building (a) the measurement operator $ \boldsymbol{C} $ is applied to the training set $ \boldsymbol{\Psi}. $ The sparse representation step (b) solves the relaxed convex optimization problem \eqref{eq:sp-approx-noise} to estimate sparse coefficients $ \boldsymbol{\hat{s}} $ which are consistent with the noisy measurements $ \boldsymbol{y}. $ Finally the full flow field is reconstructed with as a linear combination of the training examples (c). The reconstructed field shown in (c) is the actual output of the sparse representation algorithm with the noisy test data and measurements shown in (b). Flow past a cylinder is discussed further in section \ref{sec:results-cylinder}.}
	\label{fig:reconstruction-diagram}
\end{figure}

Inspired by the work of Wright et al.~\citep{Wright2009} on sparse representation for image recognition, we propose searching for a sparse representation in a library of example flow fields rather than a minimum-energy solution in the modal library, as shown schematically in figure \ref{fig:reconstruction-diagram}. 
If the flow is statistically stationary and the library is sufficiently extensive, it may be possible to identify a sparse combination of the few most similar fields in the library that are consistent with the measurements.
If such a sparse representation is available, this approach corresponds to searching in prior data for recurring coherent structures, which may be nonlinearly correlated with observations.
Moreover, sparsity-promoting techniques are known to prevent overfitting and provide robustness to noisy and corrupt measurements, which are essential for flow field estimation. 
Sparse representation has been previously applied to classify flow regimes in a library of POD modes based on limited sensors in the seminal work of~Bright, Lin, and Kutz~\citep{Bright2013}, and sparse flow classification has been extended in related work~\citep{Brunton2014,Bright2016,Kramer2017}. 
Our work builds on this framework, extending regime classification to full flow field reconstruction and demonstrating sparse representation in a library of training examples instead of a modal basis.  
We show that when the flows are more complex than previously studied, or the measurements are corrupted, this method significantly outperforms reconstruction with a POD library.  
Since highly complex flows may not have a sparse representation in terms of POD modes, this result reinforces the importance of sparsity for robustness to noise and accuracy of reconstruction.  
We extensively explore this method on several example systems of increasing complexity with various levels of measurement noise, including numerical and geophysical data sets, and find that it produces more accurate and robust reconstructions than standard least-squares methods.
Flow field estimation continues to be an important and difficult challenge, but improvements in robustness and accuracy are consistent with previous work in sparse regression and highlight the value of sparsity in reconstruction methods.

\smallskip

The remainder of this work is organized as follows.  
Section~\ref{sec:background} provides an overview of related work on flow field estimation.  
In section~\ref{sec:method}, we describe the proposed method for flow field reconstruction based on sparse representation. 
The four flow configurations used to test this method are described in detail in section~\ref{sec:configurations}.  
Section~\ref{sec:results} explores sparse reconstruction for flow field reconstruction on these examples with various sampling strategies from corrupted measurements, demonstrating that sparse representation exhibits improved robustness and accuracy compared to least-squares estimation.  
These results include careful benchmark comparisons on four fluid flows, including two canonical flows and two geophysical flows: periodic vortex shedding past a cylinder at $ \Re=100 $ (section~\ref{sec:results-cylinder}), a mixing layer at $ \Re=720 $ (section~\ref{sec:results-mixing}), global sea surface temperature fields (section~\ref{sec:results-sst}), and data from a Gulf of Mexico ocean model (section~\ref{sec:results-hycom}). 
On the mixing layer and Gulf of Mexico ocean data, we demonstrate that when a globally sparse representation is not available, the flow field can be more accurately estimated with a superposition of local reconstructions and show that this improvement is facilitated by enhanced sparsity of the local representations.  
In section \ref{sec:discussion} we summarize the main results and discuss limitations of the method, and we conclude with section~\ref{sec:conclusion}. 
To promote reproducible research, all code is available online\footnote{github.com/jcallaham/robust-flow-reconstruction}.

\section{Prior work in flow field reconstruction}\label{sec:background}
Because of its far-reaching applications, flow field estimation is a rich field with seminal works spanning the past half century.  Here, we provide a brief overview of some of the most relevant related works, which are organized into three broad groups: stochastic estimation, model-based observers, and library-based reconstruction.  Each of these methodologies approaches the problem with a slightly different motivation, but many modern studies are driven by the overarching goals of estimation and control.

Stochastic estimation (SE) was introduced by Adrian~\citep{Adrian1975} to study coherent structures in turbulence. SE is a statistical technique that estimates a quantity of interest in the flow as a conditional average given the measurements.  Expanding the conditional average in a Taylor series and minimizing the mean-square estimation error yields a functional dependence between the observation and flow field variables determined by unconditional statistics, such as the two-point correlation tensor.  SE has been extended to the estimation of POD coefficients~\citep{Bonnet1994}, spectral coefficients~\citep{Ewing1999, Tinney2006}, and inclusion of time-delayed measurements~\citep{Durgesh2010, Ukeiley2008}. The stochastic estimation method has been used to study isotropic turbulence~\citep{Adrian1975,Adrian1979, Tung1980}, turbulent boundary layers~\citep{Guezennec1989, Naguib2001}, axisymmetric jets~\citep{Bonnet1994, Tinney2006}, the backwards-facing step~\citep{Cole1998, Taylor2004, Hudy2007}, open cavities~\citep{Murray2003, Murray2007a}, and for feedback in closed-loop control of flow separation over an airfoil~\citep{Pinier2007}.  

In another approach to flow field estimation, an observer \(  \)dynamical system is used to evolve the estimate of the system state according to a reduced-order model while measurements provide feedback used to improve the estimate. The model may be linear, for example, based on dynamic mode decomposition (DMD)~\citep{Schmid2010, Rowley2009, Tu2014a, Kutz2016book}, with an estimate maintained by Kalman filtering~\citep{Tu2013, Surana2016, Kramer2017}. The model could also be nonlinear, based on a Galerkin projection of the Navier-Stokes equations onto a set of POD modes~\citep{Holmes1996,Noack2003jfm,Buffoni2008}, or the result of model identification~\citep{Loiseau2017jfm,Loiseau2018}. Recent work has investigated the use of data assimilation techniques (e.g. particle filters or ensemble Kalman filters) to estimate the mean flow~\citep{Suzuki2012jfm,Foures2014, Symon2017} or the full flow field~\citep{Combes2015, Kikuchi2015, Mons2016, daSilva2018,Silva2019phd}. In any case, the accuracy of observer-based methods depends on the quality of the reduced-order model, so there is inevitably a tradeoff between low-latency and high-accuracy models.  There are a number of excellent reviews of modal decomposition and model reduction in fluids~\citep{Rowley2017,Taira2017} .

A third category of model-free flow field estimation takes advantage of large offline data sets via library-based reconstruction.  Often the flow field will be discretized and reshaped into a high-dimensional vector, which is then approximated by a linear combination of modes in a library~\citep{Taira2017}.  The modes may be generic (e.g. a Fourier or wavelet basis) or  tailored to the particular flow (such as POD or DMD modes), which each have advantages for sensor-based flow reconstruction~\citep{Manohar2017}. 
Advanced data-driven algorithms such as K-SVD~\citep{Aharon2006, Elad2006} and GOBAL~\citep{Mathelin2018} may also be used. 
Gappy POD~\citep{Everson1995} is a popular library-based method, where the library consists of POD modes and coefficients are estimated by least-squares regression based on limited or masked data~\citep{Bui-Thanh2004, Wilcox2006, Murray2007a}.  
Podvin et al.~\citep{Podvin2005} used gappy POD to estimate POD coefficients for a 3D flow past a cavity from 2D PIV data, choosing the number of measurements to equal the number of modes, resulting in the inversion of a square matrix.  
In a related approach, Yu and Hesthaven~\citep{Yu2018} use deep learning to estimate POD coefficients. 
More generally, deep learning is a powerful emerging technique to represent multi-scale flow structure and model turbulence closure~\citep{Zhang2015aiaa,Ling2016jfm,Kutz2017,Duraisamy2018arfm}, although it generally requires tremendous amounts of training data and may be prone to overfitting unless care is taken to constrain the models with known physics.  
In a sense, deep learning may be considered a sophisticated nonlinear interpolation scheme that leverages a large library of historical examples~\citep{Mallat2016prsa}.
With different choices of library and optimization problem, the library-based reconstruction framework also encompasses compressed sensing and the sparse representation approach outlined here, although typically a library of POD modes is used with least-squares regression.

The estimation and reconstruction algorithms described above are generally based on $ \ell_2$-optimization, which suffers from the same limitations as standard least-squares parameter estimation; sparsity-promoting methods have emerged as a principled way to address these shortcomings by regularizing the regression~\citep{Xu2010,Kutz:2013,Brunton2018book,Zheng2018arxiv}.  Sparse representation in a library takes advantage of known structure in the data and is robust to measurement corruption \citep{Wright2009}.  If the coefficient vector of the modal representation is sparse in the sense that it has relatively few nonzero entries, the coefficients can be recovered from surprisingly few measurements with efficient tools, such as matching pursuit \citep{Mallat1993, Pati1993, Tropp2007, Needell2009} or by $ \ell_1 $-minimization of the coefficient vector \citep{Donoho2006a, Candes2006a}, under certain assumptions.  
If the library consists of generic modes, such as a discrete cosine transform (DCT) basis, then recovery based on $ \ell_1 $-minimization is known as compressed sensing (CS) \citep{Donoho2006, Candes2006b, Baraniuk2007}. Compressed sensing has been used in fluid mechanics to reconstruct a signal in a linear-duct acoustic problem \citep{Huang2013}, identify dominant frequencies in low-dimensional projections of sub-Nyquist rate PIV data \citep{Tu2014}, and to find a compact representation for wall-bounded turbulence \citep{Bourgiuignon2014}.  
Although these results are promising, general flows are often \emph{not sparse enough} to take advantage of compressed sensing, requiring prohibitively many measurements and expensive computations that do not scale well.   

The sparsifying library does not need to be universal, however. Sparse representation in a data-driven POD basis was used to classify the Reynolds number for flow past a cylinder~\citep{Bright2013,Bright2016}.  Bai et al.~\citep{Bai2015} similarly demonstrated CS in a POD library to reconstruct PIV data.  
Wright et al.~\citep{Wright2009} proposed a straightforward alternative to modal libraries, such as DCT and POD, in the sparse representation for classification (SRC) algorithm for facial recognition. In SRC, an image of an individual is downsampled and approximated with a sparse representation  in terms of a library consisting of \emph{the training data itself}, which contains some example images of the same person. The coefficients corresponding to the test individual will naturally be of greater magnitude, indicating the identity of the subject. 
SRC is robust to noise, corruption, or occlusion of the image, and has been applied for early diagnosis of Alzheimer's disease \citep{Liu2012}, segmentation of MRI images \citep{Tong2013}, automatic detection and classification of brain tumors \citep{Nabizadeh2015}, music genre categorization \citep{Panagakis2009}, and dolphin whistle classification \citep{Esfahanian2014}. Although SRC was introduced for classification, the success of sparse representation in a library of the training data has far reaching applications, including for flow field reconstruction, as will be explored here.  


\section{Sparse representation of a flow field in a library}\label{sec:method}
In this work, we will investigate the utility of sparse representation for flow field reconstruction in a library of historical flow field data, exploring robustness and scaling with flow complexity.  
This section provides the methodological foundations for the results that follow.  
We describe the general library-based signal recovery framework in section \ref{sec:method-general}, including reconstruction from sparse representation. In section \ref{sec:method-fluids} we introduce our method for sparse representation-based flow field reconstruction.

\subsection{Library-based signal recovery}
\label{sec:method-general}
Here we provide the general problem statement and notation for sensor-based reconstruction of a high-dimensional state in a library.  
Given a discretized state vector $ \boldsymbol{x} \in \R^n $, for example representing the fluid velocity or vorticity field at a set of grid points, and linear measurements $ \boldsymbol{y} = \boldsymbol{C}\boldsymbol{x} $, with $ \boldsymbol{y}\in\R^{p}$ and $p\ll n$, we seek an estimate $ \boldsymbol{\hat{x}} $ of the full signal.  
We assume that the state $\boldsymbol{x}$ can be accurately expressed as a linear combination of library elements $ \left\{\boldsymbol{\psi}_j\right\}, \; j=1, 2, \dots, r $ with $ \boldsymbol{\psi}_j \in \R^n $, so that
\begin{equation} \label{eq:library-reconstruction}
\boldsymbol{{x}} \approx \boldsymbol{\Psi} \boldsymbol{s},
\end{equation} 
for some coefficient vector $ \boldsymbol{s} \in \R^r, $ where columns in the library $ \boldsymbol{\Psi} \in \R^{n\times r} $ are the vectors $ \boldsymbol{\psi}_j. $ The reconstruction problem reduces to estimating the coefficients $ \boldsymbol{\hat{s}} $ that satisfy 
\begin{equation} \label{eq:square-approx}
\boldsymbol{y} \approx \boldsymbol{C} \boldsymbol{\Psi \hat{s}}.
\end{equation} 
In other words, we seek an estimate that produces measurements $\boldsymbol{\hat{y}} = \boldsymbol{C} \boldsymbol{\Psi \hat{s}}$  consistent with actual observations $\boldsymbol{y}$.  
As described earlier, the library $\boldsymbol{\Psi}$ may comprise a modal basis, such as Fourier, wavelets, POD, or DMD modes, or it may be chosen to contain examples of flow fields from training data. 
We will also explore reconstruction for different classes of measurement matrix $\boldsymbol{C}$, although it is also possible to tailor this matrix for a given library~$\boldsymbol{\Psi}$ for improved reconstruction~\citep{Manohar2017}.


In practice, solving for $ \boldsymbol{\hat{s}} $ in Eq. \eqref{eq:square-approx} must be formulated as an optimization problem, since $ \boldsymbol{C}\boldsymbol{\Psi} $ is not typically a square matrix. 
For instance, in the overdetermined case where $ p > r $ and there are more measurements than library elements, we may choose to solve for the least-squares solution 
\begin{equation} \label{eq:overdetermined-L2}
\boldsymbol{\hat{s}} = \arg \min_{\boldsymbol{s}} \| \boldsymbol{y} - \boldsymbol{C} \boldsymbol{\Psi s}\| _2.
\end{equation} 

It is generally useful to modify the least-squares regression by adding a regularization term to prevent overfitting and promote robustness to noise and outliers in the data:
\begin{equation}\label{eq:overdetemined-regularized}
\boldsymbol{\hat{s}} = \arg \min_{\boldsymbol{s}} \| \boldsymbol{y} - \boldsymbol{C} \boldsymbol{\Psi s}\| _2 + \lambda\| \boldsymbol{s}\| _q.
\end{equation} 
A choice of $ q=2 $, corresponding to Tikhonov or ridge regression, penalizes high-variance solutions. For instance, Buffoni~\citep{Buffoni2008} employed this regularization to estimate POD coefficients in a nonlinear observer. A choice of $ q=1 $ (LASSO regression) promotes a sparse representation \citep{Tibshirani1996}. The regularization parameter $ \lambda $ can be tuned to adjust the strength of this term.  Other choices of $q$ are possible, but $ q=1 $ and $ q=2 $ are the most common since they can be solved with convex optimization~\citep{Boyd2009book}, which scales well to large problems.

For high-dimensional and multiscale data, it is often the case that there are fewer available measurements than modes in the library, leading to an underdetermined problem with $ p < r $. In this case the appropriate optimization problem is 
\begin{equation} \label{eq:underdetermined-approx}
\boldsymbol{\hat{s}} = \arg \min_{\boldsymbol{s}} \| \boldsymbol{s}\| _q \hspace{1cm} \text{subject to} \hspace{1cm} \boldsymbol{y} = \boldsymbol{C} \boldsymbol{ \Psi s}.
\end{equation} 
Again, $ q=2 $ leads to the minimum-energy solution consistent with measured data, while $ q=1 $ leads to a sparse representation in the library $ \boldsymbol{\Psi}$.

If the coefficient vector $ \boldsymbol{s} $ is known to be sparse, and it is assumed to have exactly $ K $ nonzero elements (i.e., the vector $\boldsymbol{s}$ is $K$-\emph{sparse}), this problem can be formulated as 
\begin{equation}
\label{eq:pursuit-approximation}
\boldsymbol{\hat{s}}  = \arg \min_{\boldsymbol{s}} \|  \boldsymbol{y} - \boldsymbol{C}\boldsymbol{\Psi s} \| _2 \hspace{1cm} \text{subject to} \hspace{1cm} \| \boldsymbol{s}\| _0 = K,
\end{equation} 
where $\|\boldsymbol{s}\|_0$ is the number of nonzero entries of $\boldsymbol{s}$. 
Although an estimate of the sparsity $ K $ may not generally be available, there are many efficient algorithms such as OMP \citep{Mallat1993, Pati1993, Tropp2007} or CoSaMP \citep{Needell2009} that can solve this problem more efficiently than Eq.~\eqref{eq:underdetermined-approx}. 

The representation problem in Eq.~\eqref{eq:underdetermined-approx} can be applied to noisy sensor measurements $\boldsymbol{y}$. In this case, the measurements are $ \mathbf{y} = \mathbf{Cx} + \boldsymbol{\eta}, $ where $ \boldsymbol{\eta} $ is a noise vector. The optimization problem then relaxes the equality constraint: 
\begin{equation} \label{eq:sp-approx-noise}
\boldsymbol{\hat{s}} = \arg \min_{\boldsymbol{s}} \| \boldsymbol{s}\| _q \hspace{1cm} \text{subject to} \hspace{1cm} \| \boldsymbol{y} - \boldsymbol{C}\boldsymbol{\Psi s}\| _2 < \epsilon,
\end{equation}
where $ \epsilon $ is an error tolerance.  
If the measurements have independent and identically distributed Gaussian noise (i.e., $ \boldsymbol{\eta} \in \R^{p} $ with $ \eta_i \sim \mathcal{N}(0, \sigma) $), $ \epsilon $ may be chosen as a multiple of the  total noise $ \sigma \sqrt{p}. $  When noise is introduced artificially, we nondimensionalize the noise level $ \sigma $ by the RMS fluctuations of the field variable in the training set.

The relaxation in Eq.~\eqref{eq:sp-approx-noise} assumes that $ \sigma $ is relatively small compared to typical fluctuations in the measurement vector $ \boldsymbol{y}$.  Wright et al.~\citep{Wright2009} describe a method for sparse representation, with $ q=1 $, to handle sparse corruption with large amplitude, where some unknown fraction $ \rho $ of random entries in $\boldsymbol{y}$ suffer from uniformly distributed corruption over the full range of observed values.  That is, the measurement $ \boldsymbol{y} $ is now $ \boldsymbol{y} = \boldsymbol{C}\boldsymbol{x} + \boldsymbol{e}, $ where $ \boldsymbol{e} $ has $ \rho p $ nonzero entries. 
The optimization problem is extended to also identify $ \boldsymbol{e} $ by direct minimization of its $ \ell_1 $ norm. If the signal is additionally corrupted by dense low amplitude noise as in \eqref{eq:sp-approx-noise}, the problem becomes
\begin{equation} \label{eq:sp-approx-corrupt}
\boldsymbol{\hat{s}} = \arg \min_{\boldsymbol{s}, \, \boldsymbol{e}}
\|  \boldsymbol{s} \| _1 +  \| \boldsymbol{e}\| _1  \hspace{1cm} \text{subject to} \hspace{1cm}
\left|\left| \begin{bmatrix}
\boldsymbol{C}\boldsymbol{\Psi} & \boldsymbol{I}
\end{bmatrix} \begin{bmatrix}
\boldsymbol{s} \\ \boldsymbol{e}
\end{bmatrix} - \boldsymbol{y} \right|\right|_2 < \epsilon.
\end{equation}
We find, consistent with their results, that although the corruption $ \boldsymbol{e} $ must be sparse in the sense that it has enough zero entries to enable identification via minimization of its $ \ell_1 $ norm, it can actually consist of a substantial fraction of total measurements (see e.g. figure \ref{fig:cylinder-robust}). 
We demonstrate reconstruction from noisy measurements in sections \ref{sec:results-cylinder} and \ref{sec:results-mixing}, but we relax the optimization problem even when noise is not explicitly added.
Since Eq.~\eqref{eq:library-reconstruction} will generally not be exact, the relaxation $ \epsilon $ plays a similar role to the regularization parameter $ \lambda $ in Eq.~\eqref{eq:overdetemined-regularized}; larger relaxations $ \epsilon $ allow sparser estimates $ \mathbf{\hat{s}} $ which still satisfy the constraint in Eq.~\eqref{eq:sp-approx-noise}.  

As mentioned earlier, there are many choices for the library of modes, and some may be more natural depending on the application.  
Fourier modes or wavelets may be useful in audio or image compression, and empirical POD modes are often used in fluids.  In order to construct ``tailored" libraries, such as POD modes, it is necessary to have a training set $ \boldsymbol{X} \in \R^{n \times m} $ of flow fields that contains representative examples. Such a training set can be obtained from simulations or experiments. 
Other libraries may be designed to be optimal in another sense. 
For instance, the K-SVD algorithm \citep{Aharon2006, Elad2006} iteratively constructs a library in which data should have a representation with some prescribed sparsity. GOBAL \citep{Mathelin2018} is a similar method that enforces observability of the library modes.  Wright et al.~\citep{Wright2009} simply use the training set as the library, so that the columns of $ \boldsymbol{\Psi} $ are the prior observations.

This reconstruction framework enables many choices for the library and the optimization formulation.  For example, gappy POD \citep{Everson1995} solves the least-squares problem \eqref{eq:overdetermined-L2} with a library of POD modes.
To identify the high-energy structures in the flow field and ensure that the problem remains underdetermined, the library of POD modes can be truncated. This may be done automatically, for instance with the hard threshold of Gavish and Donoho~\citep{Gavish2014}, although it is not clear in general what level of truncation is optimal for flow reconstruction.

\subsection{Flow field reconstruction from sparse representation}
\label{sec:method-fluids}

\begin{figure}
	\centering
	\begin{overpic}[width=0.925\linewidth]{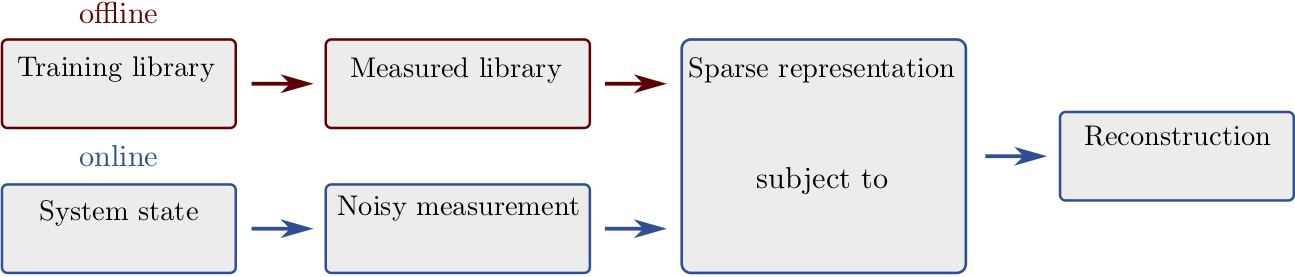}
		\put(5.6, 12.7){$ \boldsymbol{\Psi} = \boldsymbol{X} $}
		\put(33.5, 12.7){$ \boldsymbol{C} \boldsymbol{\Psi} $}
		\put(53.8, 11){$ \boldsymbol{\hat{s}} = \arg \min_{\boldsymbol{s}} \| \boldsymbol{s}\| _1 $}
		\put(54.5, 2.7){$ \| \boldsymbol{y} - \boldsymbol{C}\boldsymbol{\Psi}\boldsymbol{s}\| _2 < \epsilon $}
		\put(8.5, 1.8){$ \boldsymbol x $}
		\put(29, 1.8){$ \boldsymbol{y} = \boldsymbol{C}\boldsymbol{x} + \boldsymbol{\eta} $}
		\put(86.7, 7.2){$ \boldsymbol{ \hat{x}} =\boldsymbol{ \Psi \hat{s}} $}
	\end{overpic}
	\caption{Schematic of the sparse representation method shown graphically in figure \ref{fig:reconstruction-diagram}. After constructing the library $ \boldsymbol{C}\boldsymbol{\Psi} $ in an offline step, the sparse coefficients $ \boldsymbol{\hat{s}} $ consistent with measurements $ \boldsymbol{y} $ are estimated. The full flow field can be reconstructed as a linear combination of the training examples in $\boldsymbol{\Psi}$.}
	\label{fig:flowchart}
	\vspace{-.1in}
\end{figure}
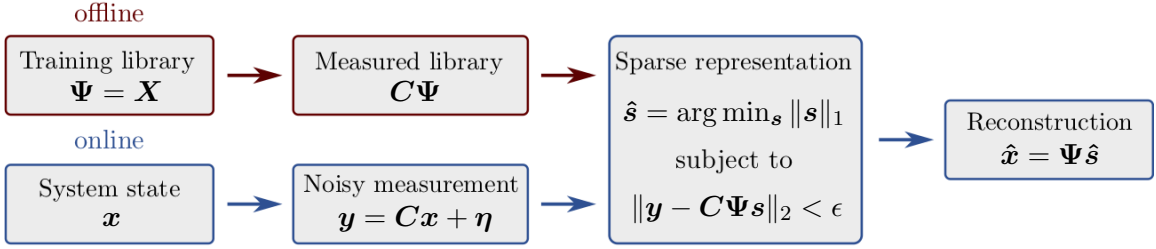

We now adapt the library-based reconstruction framework for fluid flow field reconstruction. 
The procedure is shown schematically in figures \ref{fig:reconstruction-diagram} and \ref{fig:flowchart}.  The signal $ \boldsymbol{x} \in \R^n $ is the full discretized flow field and the measurement operator $ \boldsymbol{C} $ still relates measurements $ \boldsymbol{y} $ to the full field by $ \boldsymbol{y} = \boldsymbol{C}\boldsymbol{x}$.  
Following the work of Wright et al.~\citep{Wright2009} in image analysis, we assume that the flow field has a sparse representation in a library of training examples $ \boldsymbol{\Psi}$, as opposed to POD modes.  That is, $ \boldsymbol{{x}} = \boldsymbol{\Psi s} $ for some $ \boldsymbol{s} $ with $ \| \boldsymbol{s}\| _0 = K \ll n. $  The coefficient vector $ \boldsymbol{s} $ can be estimated using one of the optimizations in Eqs.~\eqref{eq:underdetermined-approx}--\eqref{eq:sp-approx-corrupt}.

In order to improve the performance of this method for systems in fluid mechanics, we make several modifications. First, the empirical mean of the training set may be subtracted from all data. This is effective in cases where the mean represents a significant fraction of the energy in the data, for instance in sea surface temperature fields (section \ref{sec:results-sst}).
Second, the amplitudes of the reconstructed flow fields can be rescaled so that the total energy of the flow is equal to that computed from the training data. This rescaling is useful for reconstruction from noisy measurements; flow fields reconstructed with \eqref{eq:sp-approx-noise} tend to have lower amplitudes than the true field, since high levels of noise can be consistent with qualitatively accurate fields of reduced amplitude.
In this work we only use this modification in section \ref{sec:results-cylinder}.

Third, we develop a method of localized reconstruction for complex fluid flows. 
In the reconstruction process above, it is assumed that the test field is a simple linear combination of global fields in the library. 
However, for flows with coherent structures at multiple spatial scales, it may be prohibitively expensive to collect enough data to have representative examples of all likely global flow fields. 
Said another way, for multi-scale flows, it is difficult to collect enough data for the library to converge to a statistical stationary distribution.  
Fortunately, if we decompose the global domain into local patches, each patch may be much lower rank, enabling a local sparse representation. 
To facilitate localized reconstruction, we introduce local kernels $ \boldsymbol{\Phi}_j, \; j=1, 2, \dots, k $ that restrict the measurement $ \boldsymbol{y}_j = \boldsymbol{C}\boldsymbol{\Phi}_j\boldsymbol{x}$ and reconstruction to the $ j- $th local region:\begin{equation}
\label{eq:kernel-reconstruction}
\boldsymbol{\hat{s}}_j = \arg \min_{\boldsymbol{s}_j} \| \boldsymbol{s}_j\| _q \hspace{1cm} \text{subject to} \hspace{1cm} \| \boldsymbol{y}_j - \boldsymbol{C}\boldsymbol{ \Phi}_j \boldsymbol{\Psi s}_j\| _2 < \epsilon.
\end{equation} These decoupled optimization problems lead to compact local estimates $ \boldsymbol{\hat{x}}_j = \boldsymbol{\Phi}_j\boldsymbol{\Psi \hat{s}}_j, $ with a full state estimate, $ \boldsymbol{\hat{x}} = \sum_j \boldsymbol{\hat{x}}_j$, that is globally valid.

Finally, we define a metric to compare the quality of various reconstructions.The normalized root-mean-square residual of the difference of the reconstruction $ \boldsymbol{\hat{x}} $ and the test field $\boldsymbol{x}$ is:
\begin{equation}
\label{eq:error-metric}
\text{error} = \frac{\| \boldsymbol{x}-\boldsymbol{\hat{x}}\| _2}{\| \boldsymbol{x}\| _2}.
\end{equation} 
If the empirical mean $ \boldsymbol{\bar{x}}$ is subtracted from both $\boldsymbol{x}$ and $\boldsymbol{\hat{x}}$, then the appropriate metric is 
\begin{equation}
\text{error} = \frac{\| \boldsymbol{x}-\boldsymbol{\hat{x}}\| _2}
{\| \boldsymbol{x}+\boldsymbol{\bar{x}}\| _2}.
\end{equation}

At this point, it is important to summarize some of the main assumptions that sparse representation relies on. First, as with all reconstruction methods based on a tailored library, we assume that the flow is statistically stationary, so that a sufficiently large library based on training data can generalize to future states. 
We then assume that the training set is comprehensive enough that the observed states are well-approximated by a linear combination of library elements.  
This is a related requirement, but while the former is a property of the flow, the latter is a property of the training data itself.  
All reconstruction methods further rely on the measurements containing sufficient information to accurately identify the coefficients $ \boldsymbol{\hat{s}}$. 
Just as reconstruction is impossible, even with relatively dense sampling, for a flow state that is essentially orthogonal to the library, we cannot realistically expect to accurately reconstruct a turbulent channel flow from one point measurement, no matter how extensive the library.
Finally, sparse representation assumes that a flow field of interest may be expressed as a linear combination of a small number of other flow fields in the training library. This assumption holds for simple flows, such as periodic vortex shedding at low Reynolds number, but may be unjustified for complex, multiscale flows unless a staggering amount of data is available. 
We examine the implications of these assumptions in the following sections.

\subsubsection{Algorithm}
The complete flow field reconstruction based on sparse representation is as follows:
\begin{enumerate}
	\item Compute the library $ \boldsymbol{\Psi}\in \R^{n \times r}$.  
The library may be given by the unmodified training data $ \boldsymbol{X} \in \R^{n \times m}, $ although we also investigate reconstruction using POD modes and a K-SVD library. Optionally, the empirical mean flow field $ \boldsymbol{\bar{x}} \in \R^n$ may be subtracted from $ \boldsymbol{X} $.
	\item Take measurements $ \boldsymbol{y} = \boldsymbol{C}\boldsymbol{x} + \boldsymbol{\eta} $ of the flow field $ \boldsymbol{x} $ using the measurement operator $ \boldsymbol{C} \in \R^{p \times n}$ with noise $\boldsymbol{\eta}$.  For the examples below, the measurement matrix $ \boldsymbol{C} $ consists of rows of the identity matrix corresponding to measured locations in the discretized field.
	\item Solve the appropriate optimization problem in Eqs.~\eqref{eq:underdetermined-approx}--\eqref{eq:sp-approx-corrupt} for the coefficient vector $ \boldsymbol{\hat{s}}. $ If the coefficients are found by minimizing the $ \ell_1- $norm, we refer to this as sparse representation.
	\item Reconstruct the estimated flow field with $ \boldsymbol{\hat{x}} = \boldsymbol{\Psi \hat{s}}. $ Optionally, rescale the estimated field to have the same variance as the training fields; this is helpful for very noisy measurements.
\end{enumerate}

For dictionary learning with K-SVD, we use KSVD-Box v13.  
We solve the pursuit problem in Eq.~\eqref{eq:pursuit-approximation} with OMP-Box v10~\citep{Rubenstein2008}. To solve the convex optimization problems \eqref{eq:sp-approx-noise}, \eqref{eq:sp-approx-corrupt}, and \eqref{eq:kernel-reconstruction}, we use the CVX Matlab package \citep{Grant2008, Grant2013}. The complexity of sparse approximation grows with the number of measurements and the number of modes in the dictionary, but does not directly depend on the number of points in the original discretized field.


\section{Flow configurations}\label{sec:configurations}

\begin{figure}
	\centering
	\begin{overpic}[width=0.925\linewidth]{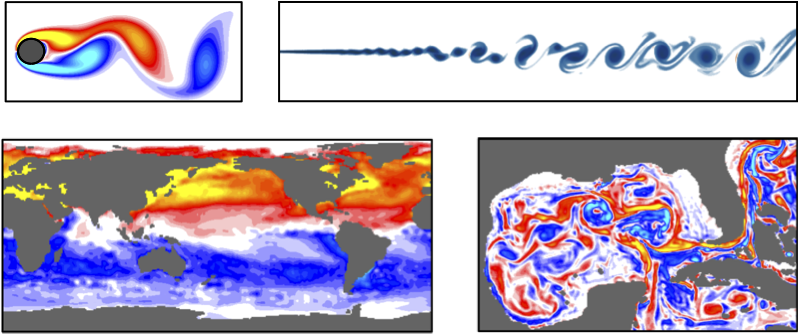}
		\put(-1, 42.5){\secref{sec:results-cylinder}: $ \Re=100 $ vortex shedding}
		\put(54, 42.5){\secref{sec:results-mixing}: $ \Re=720 $ mixing layer}
		\put(12, 25.5){\secref{sec:results-sst}: Sea surface temperature}
		\put(64, 25.5){\secref{sec:results-hycom}: Gulf of Mexico vorticity}
	\end{overpic}
	\vspace{-.1in}
	\caption{Example flow fields from the data sets which we investigate with the sparse representation-based reconstruction method. We study two canonical flows (periodic vortex shedding past a cylinder at $ \Re=100 $ and a mixing layer at $ \Re=720 $) and two geophysical data sets (sea surface temperature and Gulf of Mexico vorticity fields). }
	\label{fig:flow-configs}
\end{figure}
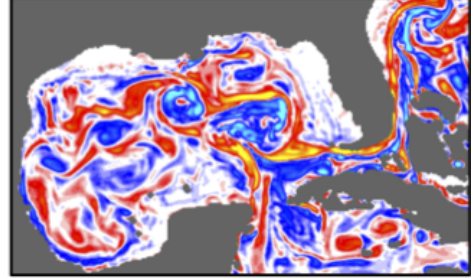

Here we describe the fluid flows explored in this work and the methods used to obtain the data.  We apply flow field reconstruction  to four data sets of increasing complexity, shown in figure~\ref{fig:flow-configs}: vortex shedding past a cylinder at $ \Re=100 $, a mixing layer at $ \Re=720 $, observations of sea surface temperature, and sea surface vorticity in the Gulf of Mexico.  

\subsection{Periodic vortex shedding}
The first test case is given by the two-dimensional fluid flow past a circular cylinder at Reynolds number $100$, which is characterized by periodic, laminar vortex shedding. 
This flow is a canonical benchmark system, although it is considerably simpler than most flows of practical interest.  

Our data was generated from direct numerical simulation of the incompressible Navier-Stokes equations using the immersed boundary projection method~\citep{Taira2007, Colonius2008}\footnote{The fluid solver used to generate this data is publicly available at https://github.com/crowley/ibpm. The data is available at www.dmdbook.com.}.  The computational domain consists of four nested grids with the smallest grid covering a domain of $9\times 4$ cylinder diameters and the largest grid covering a domain of $72 \times 32 $ diameters.  
The resolution for each grid is $ 450\times200 $ (50 points per cylinder diameter) and the simulation uses a time step of $ \Delta t = 0.02 $ time units that are non-dimensionalized by the free-stream velocity and the cylinder diameter. 
We collect 151 post-transient snapshots, corresponding to five periods of vortex shedding, with each snapshot separated by $ 10\Delta t$. 
The Reynolds number for this flow, based on the cylinder diameter and free-stream velocity, is $ \Re=D U_\infty/\nu=100$, where $ D $ is the diameter, $ U_\infty $ is the free-stream velocity, and $ \nu $ is the kinematic viscosity. 
The training set consists of the first 32 snapshots, which spans one full period.  
We analyze the vorticity field, although the method could be applied to velocity, pressure, scalar concentrations, or any other field variables of interest.  
The mean vorticity field is included in visualizations, although analyses and error calculations are performed after subtracting the empirical mean of the training data. 

\subsection{Mixing layer}
As a more complex example, we consider a two-dimensional, compressible mixing layer at Reynolds number 
\begin{equation*}
\Re = \frac{\Delta U \, \delta }{\nu} = 720,
\end{equation*} 
where $ \delta $ is the initial vorticity thickness, $\Delta U $ is the velocity difference across the layer, and $ \nu $ is the kinematic viscosity. Stanley and Sarkar~\citep{Stanley1997} showed that two-dimensional numerical simulations in this regime reproduce the flow structures observed in three-dimensional experiments.

We generated this data set by direct numerical simulation of the compressible Navier-Stokes equations using a finite volume, 5th-order WENO scheme \citep{Coralic2014}.  The spatial coordinates are normalized by the vorticity thickness at the inlet. The velocities are normalized by the speed of sound of the fluid far from the mixing region. The Mach numbers of the high- and low- stream velocity are 0.5 and 0.25, respectively.  The computational domain is $ x\in[0,800] $ and $ y\in[-200,200] $.  The flow is forced at the inflow boundary at its most unstable fundamental frequency, and its sub-harmonic. Non-reflective boundary conditions are implemented on the other boundaries. The grid is smoothly stretched away from the mixing region to the non-reflective boundaries to prevent contamination by reflections. The grid in the mixing region is uniform with $ \Delta x $ = 0.08 and $ \Delta y = 0.02 $, respectively.  After removing the transient portion of the simulation, we collect 2400 snapshots of the vorticity field in a window of $ (x, y) \in (0, 128)\times(-12, 12) $, separated by non-dimensional time steps of $ \Delta t = 0.5827. $ We compute the normal vorticity from these velocity fields both for ease of visualization and for the importance of vorticity in identifying dynamically significant coherent structures \citep{Hussain1981}.

Forcing at the inlet excites instability waves, which roll up into vortices and convect downstream. 
 These vortices pair and eventually merge into successively larger vortices. 
 This process contributes to the linear growth of the mixing layer~\citep{Winant1974}, and at higher Reynolds number, the turbulent mixing layer is dominated by the linear growth of the coherent structures~\citep{Brown1974}. These structures play a significant role in mixing, transport, and entrainment in turbulent shear flows~\citep{Hussain1981}.  
 Therefore, we study this laminar mixing layer as a representative case to assess the ability of sparse representation to generalize to other shear flow configurations.


\subsection{Sea surface temperature field}
Real-world data is rarely as well-behaved as the numerical solutions of canonical flows. 
However, it is these flows, with limited training data, multiscale dynamics, and unmodeled coupling to external systems, where the ability to infer the structure of the field would be most useful. To this end we explore reconstruction methods with the NOAA Optimum Interpolation Sea Surface Temperature (SST) V2 data set\footnote{NOAA OISST v2 data provided by the NOAA/OAR/ESRL PSD, Boulder, Colorado, USA, from https://www.esrl.noaa.gov/psd/ (Accessed 2018).}.  Due to seasonal fluctuations, the SST field exhibits strongly periodic structure, although complex ocean dynamics still lead to rich flow phenomena.  
Flow data is available on a weekly basis on a one degree grid, and it is produced by combining local and satellite temperature observations.  
We use all available data at the time of analysis (1914 weeks, spanning October 1981-June 2018). Our training set consists of the first 20 years of data (1040 weeks, spanning 1981-2001).  
We calculate a long-term annual mean field and subtract the mean for all analyses, since the mean accounts for the majority of the spatial structure of the field and is therefore uninformative with respect to the performance of reconstruction methods.  

\subsection{Gulf of Mexico surface vorticity}
Finally, we consider the Gulf of Mexico surface velocity estimates from the HYbrid Coordinate Ocean Model (HYCOM) group. This data-assimilative model synthesizes remotely sensed and in situ measurements on a hybrid coordinate system\footnote{Funding for the development of HYCOM was provided by the National Ocean Partnership Program and the Office of Naval Research. Data assimilative products using HYCOM are funded by the U.S. Navy. Computer time was made available by the DoD High Performance Computing Modernization Program. Data available at http://hycom.org.}. We use daily $ 1/12.5^\circ$-resolution data from 1992-2018 (9268 snapshots, combined from HYCOM experiment numbers 19.0, 19.1, 90.9, 91.1, and 91.2).  The training set consists of 8341 snapshots, or approximately 90\% of the total data, with the remainder withheld for independent validation.  As with the mixing layer and cylinder, we compute vorticity from 2D velocity measurements, although the methods are readily applied to any quantity of interest.  We analyze the fluctuating vorticity fields relative to the empirical mean of the training set, but include the mean flow in visualizations.

\begin{figure}
\vspace{-.25in}
	\centering
	\begin{overpic}[width=0.525\linewidth]{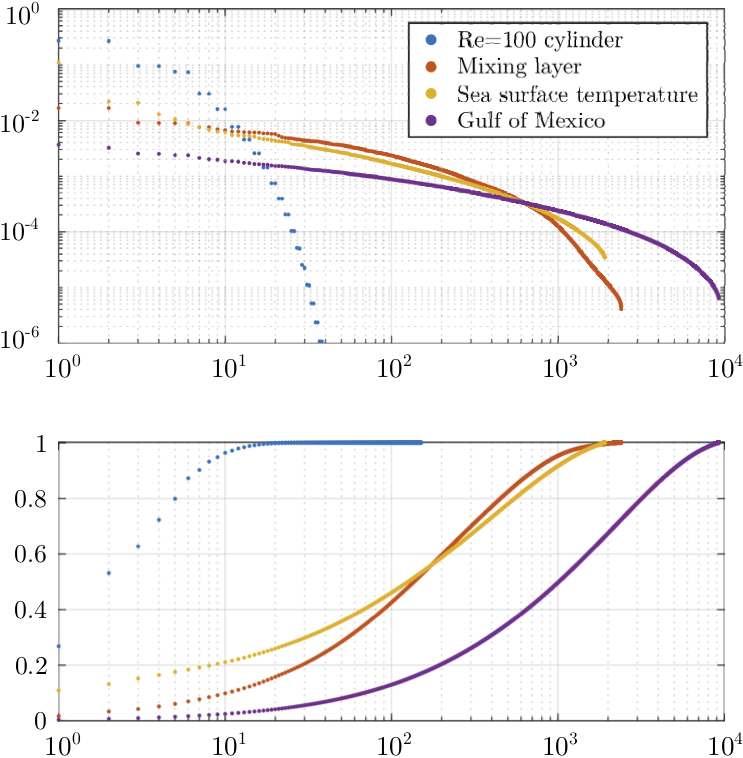}
		\put(39.5,-4.5){Mode number}
		\put(-4.5,8){\begin{sideways}Cumulative energy\end{sideways}}
		\put(-4.5,51.5){\begin{sideways}Normalized singular value\end{sideways}}
	\end{overpic}
	\vspace{.1in}
	\caption{Singular value spectra for the flows studied in this work. Each is normalized by the sum of singular values for that flow. The normalized cumulative sum of the singular values (bottom) represents the energy captured in the dominant POD modes. The rate of convergence gives an indication of the complexity of the flow. The singular values for vortex shedding past a cylinder (blue) converge quickly, whereas the Gulf of Mexico vorticity data (purple) has a long tail. The sea surface temperature (yellow) and mixing layer vorticity (red) are of intermediate complexity.}
	\label{fig:sv-spectra}
	\vspace{-.1in}
\end{figure}

Figure \ref{fig:sv-spectra} shows the singular value spectra, equivalent to the POD eigenspectra, of the four data sets.  This offers a rough comparison of the complexity of the flows. The low-dimensional dynamics of the flow behind a cylinder is clear from the sharp decay of singular values; most of the energy is contained within the first twenty POD modes.  On the other hand, the spectrum for the Gulf of Mexico data converges slowly, indicating complex multiscale dynamics.  
The difficulties with this data set are intuitive: by restricting our view to the Gulf of Mexico we study a flow with an unmodeled coupling to a much larger chaotic system. 
The mixing layer and sea surface temperature data exhibit intermediate complexity.  For the mixing layer, the flow near the inlet is approximately periodic, and the behavior becomes more complex as the flow evolves downstream.  Similarly, the SST fields show strong seasonal fluctuations with perturbations.  

\section{Results}\label{sec:results}

We now investigate flow field reconstruction from limited measurements using the four data sets described in section \ref{sec:configurations}, which span a range of physical scales and complexity.  
For the two numerically generated flows, we study the impact of measurement noise on reconstruction accuracy and find improved robustness with sparse representation.  
In the mixing layer, we demonstrate the advantages of sparse representation by analyzing the flow in windows to enable higher levels of sparsity.  
Finally, we demonstrate the proposed method on two geophysical data sets: global sea surface temperature and Gulf of Mexico surface vorticity.  In all cases, we compare the performance of sparse representation to other library-based methods, including gappy POD.

\subsection{Periodic vortex shedding}\label{sec:results-cylinder}

\begin{figure}
	\centering
	\hspace{.05in}
	\begin{overpic}[width=0.925\linewidth]{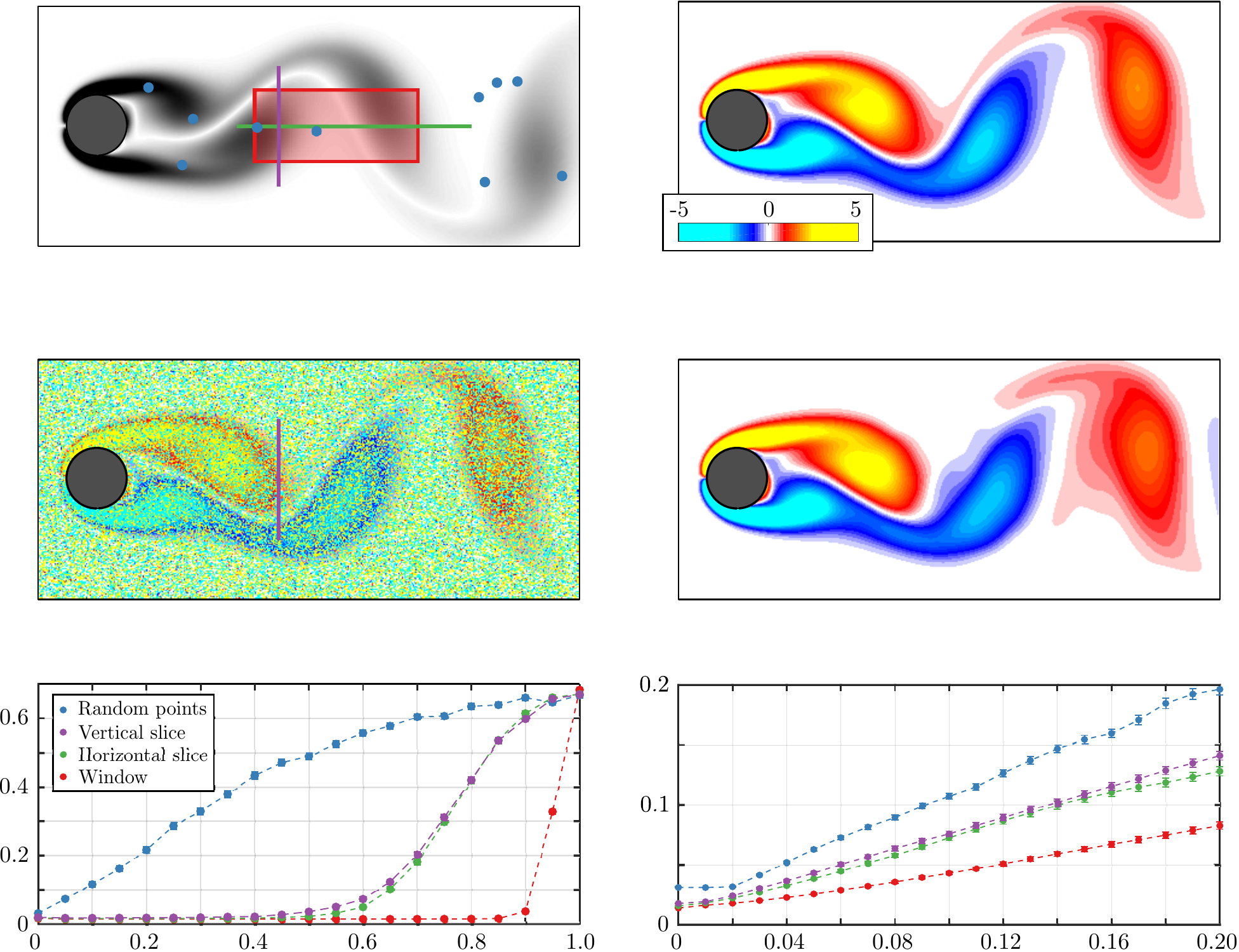}
		\put(3, 77.5){a)}
		\put(13, 77.5){Measurement strategies}
		\put(55, 77.5){b)}
		\put(71, 77.5){Test field}

		\put(3, 49.){c)}
		\put(14, 49.){Corrupt measurement}
		\put(-2.5,0){\begin{sideways}Reconstruction error\end{sideways}}\
		\put(14, -3.){Percent corruption $\rho$}

		\put(55, 49.){d)}
		\put(66.5, 49.){Reconstructed field}
		\put(49.5,0){\begin{sideways}Reconstruction error\end{sideways}}
		\put(68,-3.){Noise level $\sigma$}
		
		\put(3, 22.5){e)}
		\put(14.5, 22.5){Error vs. corruption}

		\put(55, 22.5){f)}
		\put(67.5,22.5){Error vs. noise}
	\end{overpic}
	\vspace{.25in}
	\caption{Sparse reconstruction in a library of training data accurately recovers the flow past a cylinder in the presence of noise and corruption using a variety of measurement strategies. (a) Illustration of different measurements: random points (blue), vertical and horizontal slices (resp. purple and green), and a window (red). (b) Flow field from the test data set. 
	(c) Example flow snapshot with corruption in 70\% of grid locations. The vertical stripe shows the measurement location. (d) Reconstructed flow field from sparse representation using corrupted measurements shown in (c). 	
	(e) Normalized reconstruction error with increasing percentage of grossly corrupted grid points (see section \ref{sec:method-general}).  Colors correspond to the measurements in (a) and error bars show standard error as obtained by simulating $10$ different realizations of the Gaussian noise for the $119$ test fields. 
	(f) Normalized reconstruction error with increasing levels of dense, normally distributed noise. 
	 In all cases, more measurements result in better performance.
	}
	\label{fig:cylinder-robust}
\end{figure}

Figure \ref{fig:cylinder-robust} demonstrates reconstruction of the flow past a cylinder from various measurements with increasing levels of noise and corruption.   
In particular, we consider sparse representation in a library of the training data with and find that this flow can be accurately reconstructed, even in the presence of significant noise. 
We also investigate various measurement strategies, including random point measurements, a ``window" inspired by PIV-type measurement, and continuous ``slices" in both vertical and horizontal orientations. 
For example, figure \ref{fig:cylinder-robust} demonstrates recovery of the entire field using Eq. \eqref{eq:sp-approx-corrupt} from a cross-stream slice measurement where 70\% of the measured points are corrupted by replacing the observed value with a uniform random value on the range of observed vorticity.  
In all cases, the measurement strategies with larger numbers of observations (e.g., the red window) exhibit more robust reconstruction performance.  
In addition, for corrupt measurements, there is a \emph{phase change} observed at a critical corruption density, consistent with the wider sparse representation literature.   
It is not surprising that sparse representation is effective for this simple example, since the flow is periodic and patterns observed in the training data generalize to the test fields. 
In fact, for this reason we do not have a truly independent set of flow fields on which we validate the ability of these methods to generalize beyond training data.
However, these results are encouraging, as sparse representation exhibits accurate and robust reconstruction across a variety of physical measurement configurations and noise intensity.  

\begin{figure}
	\centering
	\vspace{0.25in}
	\begin{overpic}[width=0.75\linewidth]{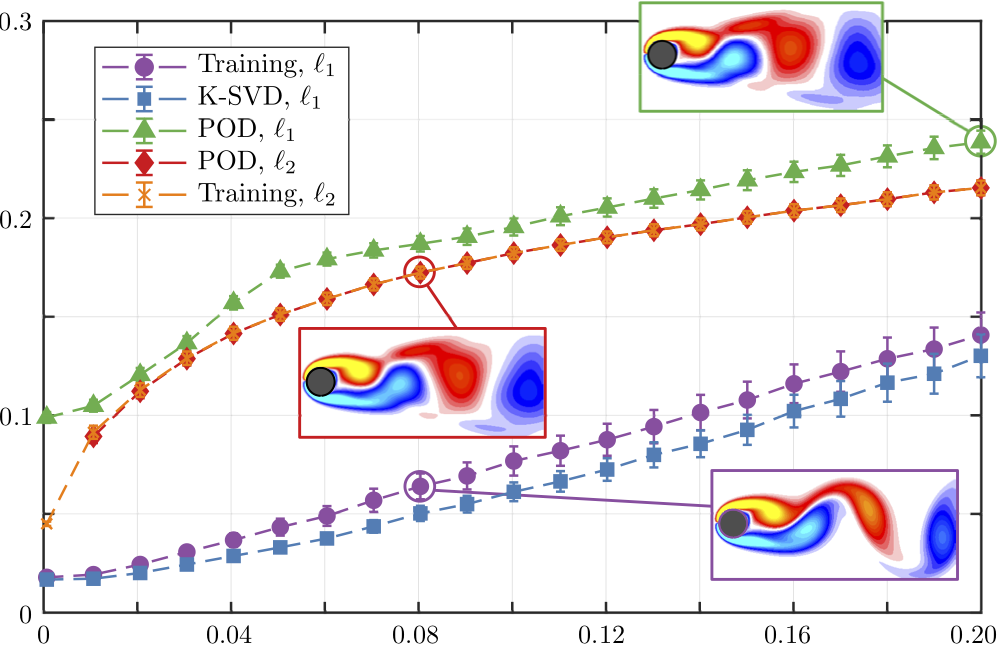}
		\put(43,-4){Noise level, $\sigma$}
		\put(-3.5,18){\begin{sideways}Reconstruction error\end{sideways}}
	\end{overpic}
	\vspace{.15in}
	\caption{Comparison of reconstruction with different libraries and norms from a noisy vertical measurement slice (figure \ref{fig:cylinder-robust}, purple). The horizontal axis is the level of dense Gaussian noise, the vertical axis is the normalized residual error in the reconstruction \eqref{eq:error-metric}, and the error bars indicate standard error on the mean residual. Insets show typical examples of reconstruction. Over this range of noise, sparse representation with the training library (red) shows an average 35\% improvement over the more standard gappy POD (purple).}
	\label{fig:cylinder-dictionaries}
\end{figure}

With the wealth of potential reconstruction techniques within the library-based optimization framework, it is interesting to explore the relative resilience to noise of different choices of the library and regularizing norm.  
Figure \ref{fig:cylinder-dictionaries} shows a comparison of reconstruction accuracy with increasing noise level for several of these combinations. 
We find that over a wide range of Gaussian noise levels, the sparse reconstruction ($ \ell_1 $ norm) with either the training library or a K-SVD library outperforms POD-based methods. 
The slope of the $\ell_2$-based methods are smaller than those of the $\ell_1$-based methods, so they will eventually achieve lower relative error, although at such large levels of noise it is unlikely that any method will result in a useful reconstruction.  
The poor performance of $ \ell_1 $ optimization with a POD library indicates that the POD basis does not admit a sparse representation; empirical POD modes are eigenfunctions of a time-averaged correlation matrix, so that the energy in any particular flow field is distributed across modes. 
Thus, although the POD basis is optimal in the sense that it offers the best global reconstruction for a given number, this does not  translate into optimality for the problem of reconstruction from limited measurements. 
In contrast, the K-SVD library is designed to admit a sparse representation of the data, and it is not surprising that this library results in the best performance\footnote{K-SVD allows for tuning several parameters; although our chosen values work well, these are not necessarily optimal.}.  
However, sparse representation in a library of the training data exhibits similar performance and benefits from simple implementation and interpretable results.  
These results reinforce the importance of sparse representation with respect to robustness to noise, a quality which has made $ \ell_1-$regularization a popular tool to prevent overfitting in parameter estimation \citep{Xu2010}.

It is not surprising that sparse representation is so effective on this example, since the flow is low-rank, periodic, and does not exhibit multiscale phenomena.  
For a periodic flow, sparse representation reduces to choosing the single example with the correct phase from the library.  
In contract, each flow field is a dense linear combination of POD modes, so that this library does not admit a sparse representation.  
In this sense, a highly sparse representation can be expected to reproduce realistic flow physics.  
Reconstruction from limited, noisy measurements via sparse representation in a training library can therefore provide a robust alternative to $ \ell_2- $based methods.

\subsection{Mixing layer}\label{sec:results-mixing}
The downstream evolution of the mixing layer leads to globally aperiodic dynamics, so we cannot expect to exactly reproduce an arbitrary flow field with a single example from in the training library, as was the case for the periodic vortex shedding behind a cylinder.  
However, we find that highly sparse representations still lead to accurate flow field estimates. 
This suggest that the library of mixing flow fields generalize to new flows that are not in the training set, allowing a sparse representation.  

\begin{figure}
	\centering
	\begin{overpic}[width=0.925\linewidth]{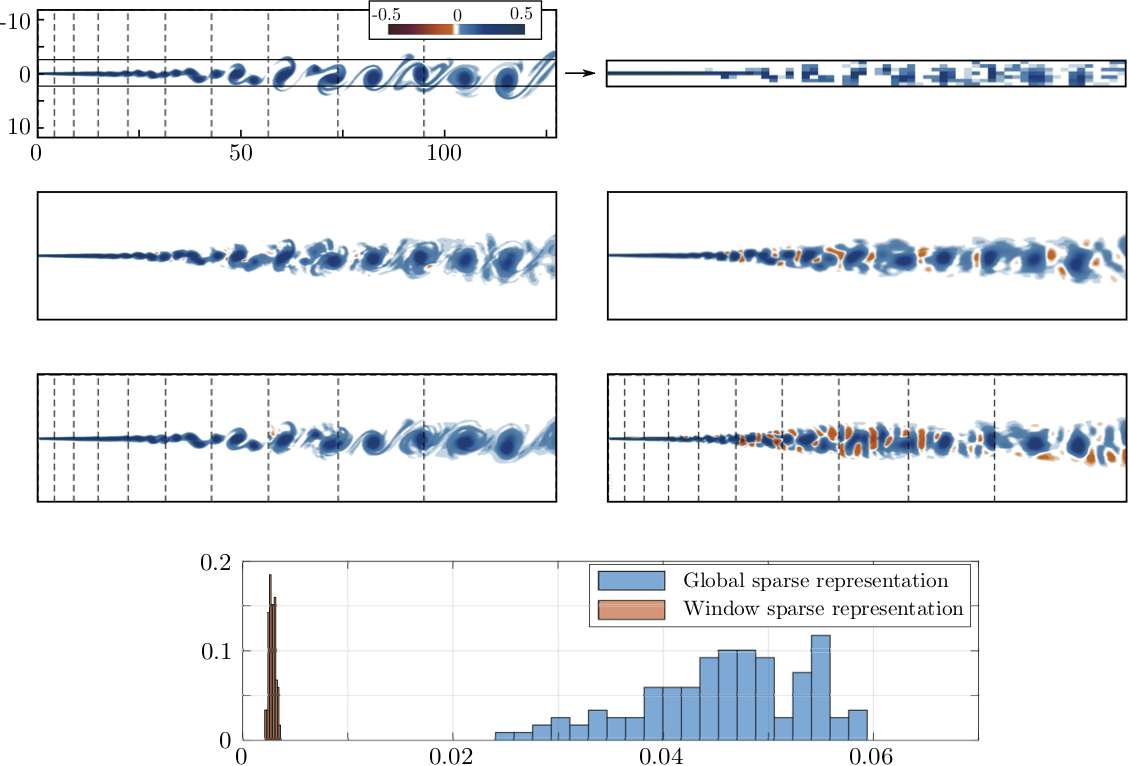}
		\put(3.3, 68){a)}
		\put(22, 68){Test field}
		\put(53.2, 68){b)}
		\put(68, 68){0.2\% sampling}
		\put(3.3, 51.4){c)}
		\put(11, 51.4){Sparse representation (global)}
		\put(53.2, 51.4){d)}
		\put(66, 51.4){Gappy POD (global)}
		\put(3.3, 35.7){e)} 
		\put(10, 35.7){Sparse representation (window)}
		\put(53.2, 35.7){f)} 
		\put(64.5, 35.7){Gappy POD (window)}
		\put(22, 19){g)}
		\put(36, 19){Relative sparsity of representation}
		\put(14.3, 5){\begin{sideways}Frequency\end{sideways}}
		\put(45, -2.6){Relative sparsity}
	\end{overpic}
	\vspace{0.15in}
	\caption{Global reconstruction of a mixing layer vorticity field via super-resolution. The test field (a) is measured with 10:1 downsampling (b) and reconstructed with both sparse representation in a training library (c) and least-squares regression in a POD library (d). Though both methods overfit the data, sparse representation captures the large-scale structures more effectively than least-squares POD.  Separating the domain into windows that grow linearly in the streamwise direction, consistent with the flow dynamics, leads to a more realistic reconstruction from sparse representation (e), although gappy POD does not improve with this approach. The relative sparsity, given by the fraction of nonzero coefficients, of the global and windowed sparse representations (g) shows that windowing enables improved sparsity.}
	\label{fig:mixing-downsample}
\end{figure}
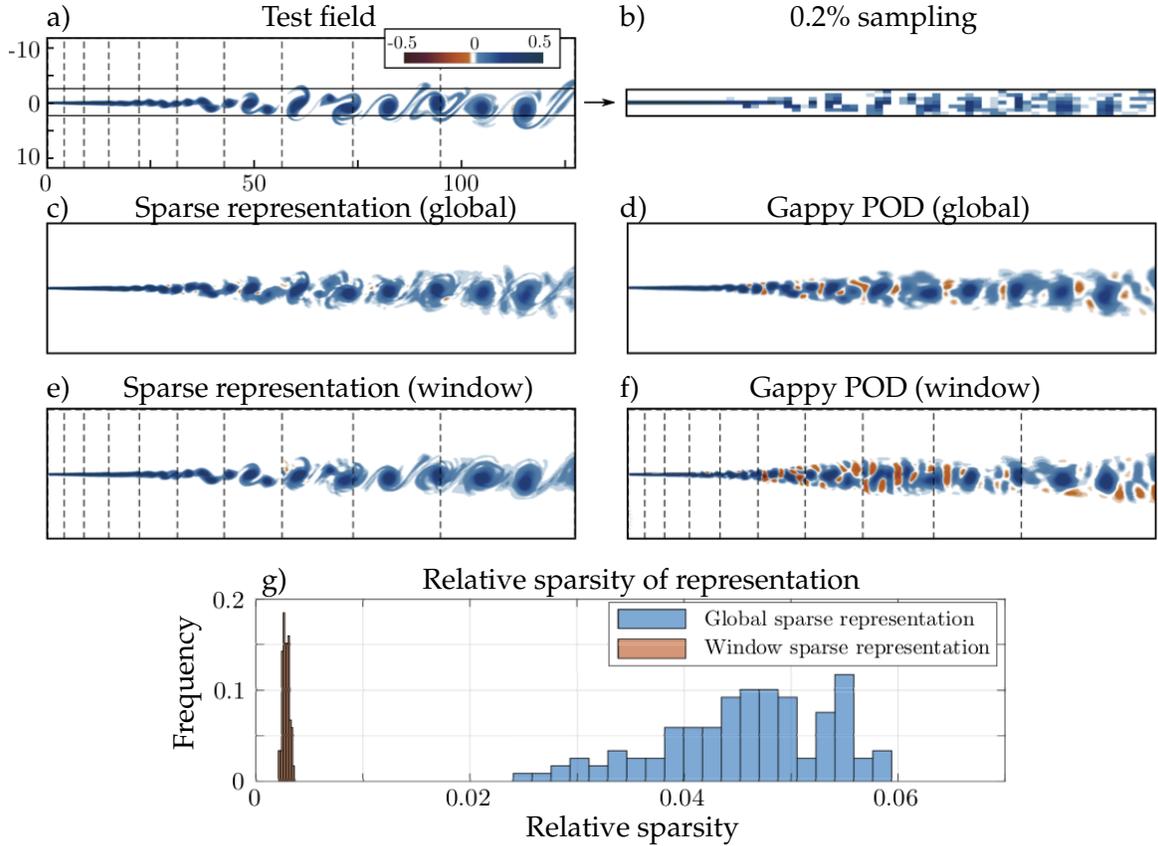

Figure \ref{fig:mixing-downsample} demonstrates reconstruction of the normal vorticity of the mixing layer from spatially downsampled measurements, given by 10:1 downsampling of the original data in the middle region containing the mixing layer.  Since noise is added to the measurements, the sparse representation is found using Eq. \eqref{eq:sp-approx-noise} with an optimally chosen value of $ \epsilon $ (see Appendix B).
This may be thought of as a \emph{super-resolution} problem~\citep{Yang2010ieeetip,Freeman2002ieeecga}, where low-resolution measurement data is synthesized into a higher-resolution field based on a high-resolution library.  
We compare reconstruction via sparse representation in the training library to $ \ell_2 $ minimization in a truncated library of POD modes ($ r= 50$).
'Both suffer from some degree of overfitting, since the specific global arrangement of vortices in the test data is likely not observed in the training data. 
Still, sparse representation builds a reasonable picture of the early perturbations and later large-scale vortical structure, whereas both are barely identifiable in the POD reconstruction.

The relatively limited accuracy of sparse representation in this case suggests that the global test field is not a sparse linear combination of examples in the training set, presumably because all possible arrangements of vortices and their phases have not been observed.  
Compared to vortex shedding past a cylinder, this flow exhibits more complex, multiscale dynamics that are driven by the successive vortex pairing process.  
The local measurements may not be informative or correlated with the global structure of the flow field, which is an implicit assumption of the global library-based estimation.  In such cases, where the global domain is larger than the de-correlation length, it may be helpful to assume that measurements inform only the local spatial region of the flow.
Thus, we apply the localized reconstruction process introduced in section \ref{sec:method-fluids}.  We divide the full flow field into ten windows that grow linearly in the streamwise direction, consistent with the streamwise dynamical scaling, and solve the local sparse representation problems independently. Reconstructions are then formed from sparse combinations of the windowed training fields. This localized method (figure \ref{fig:mixing-downsample}e) outperforms the global reconstruction from sparse representation (figure \ref{fig:mixing-downsample}c), and both significantly improve upon POD-based reconstructions (figure \ref{fig:mixing-downsample}d, f).

\begin{figure}
	\centering
	\begin{overpic}[width=0.925\linewidth]{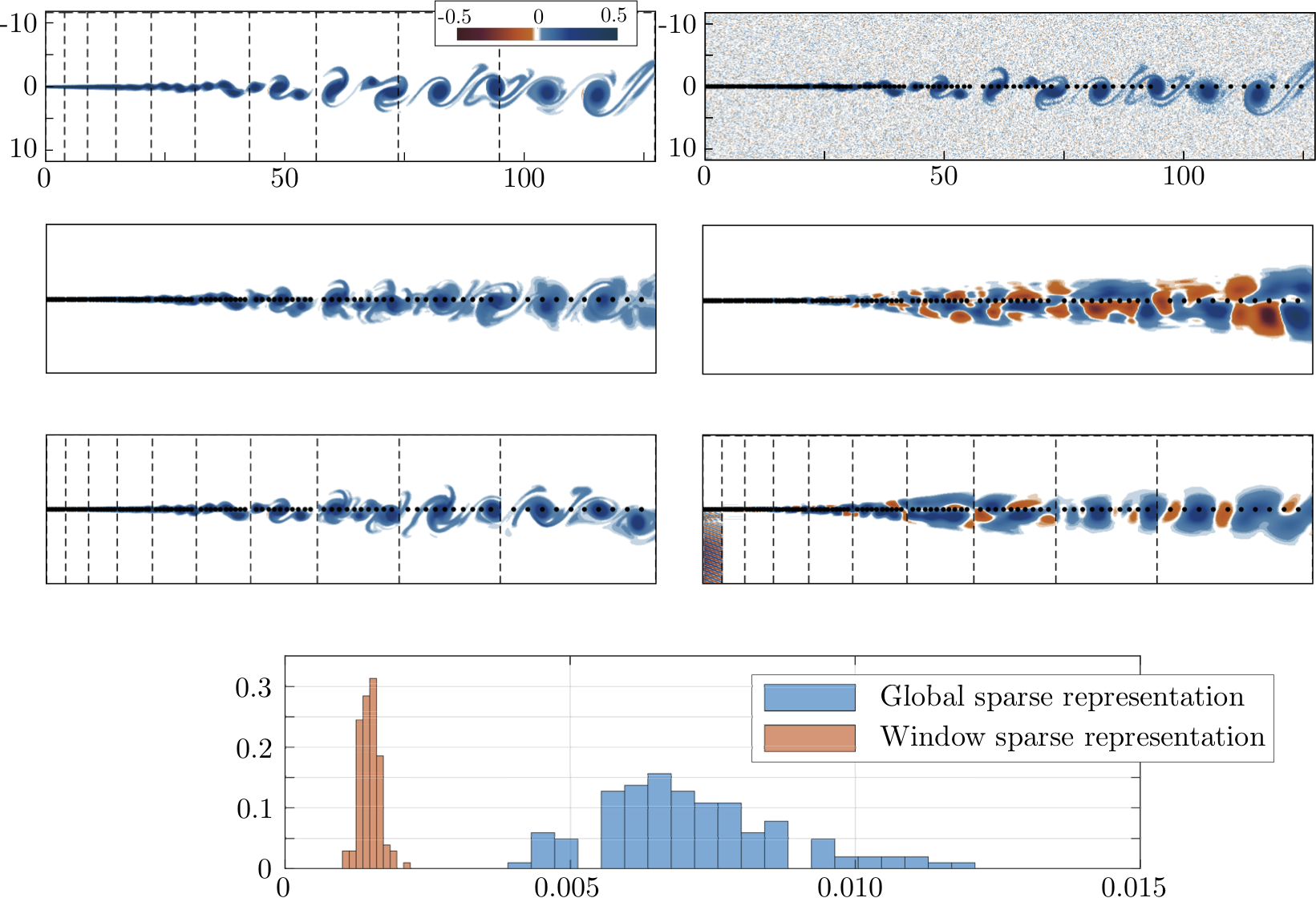}
		\put(3.5, 68.2){a)}
		\put(22, 68.2){Test field}
		\put(53.2, 68.2){b)}
		\put(68, 68.2){Noisy measurement}
		\put(3.5, 51.8){c)}
		\put(11, 51.8){Sparse representation (global)}
		\put(53.2, 51.8){d)}
		\put(66, 51.8){Gappy POD (global)}
		\put(3.5, 36){e)} 
		\put(10, 36){Sparse representation (window)}
		\put(53.2, 36){f)} 
		\put(64.5, 36){Gappy POD (window)}
		\put(22, 19){g)}
		\put(36, 19){Relative sparsity of representation}
		\put(15, 5){\begin{sideways}Frequency\end{sideways}}
		\put(45, -2.6){Relative sparsity}
	\end{overpic}
	\vspace{0.2in}
	\caption{Local mixing layer vorticity field reconstruction.  As in figure~\ref{fig:mixing-downsample}, we construct windows that grow linearly in the spanwise direction and collect ten point measurements at noise level $ \sigma =  0.3$ (white and black dots) from the mixing layer centerline in each window (separated by dashed lines).  Local reconstructions are computed based on a windowed library of training examples (e, f).  The normalized residual errors for these examples are 0.50 for the global sparse reconstruction (c), 0.40 for the windowed sparse reconstruction (e), 1.29 for global gappy POD reconstruction (d), and 0.98 for the local POD estimate (f).  Although the windowed sparse representation has larger global errors than the global estimation, the local reconstruction is more accurate in the windows closer to the inlet, since the time scales of the flow are shorter there and the training set is more likely to contain examples of similar fields (see also figure \ref{fig:pod-residual}b).
	The relative sparsity of the global and windowed sparse representations in (g) shows improved sparsity with windowing. }
	\label{fig:mixing-centerline}
\end{figure}
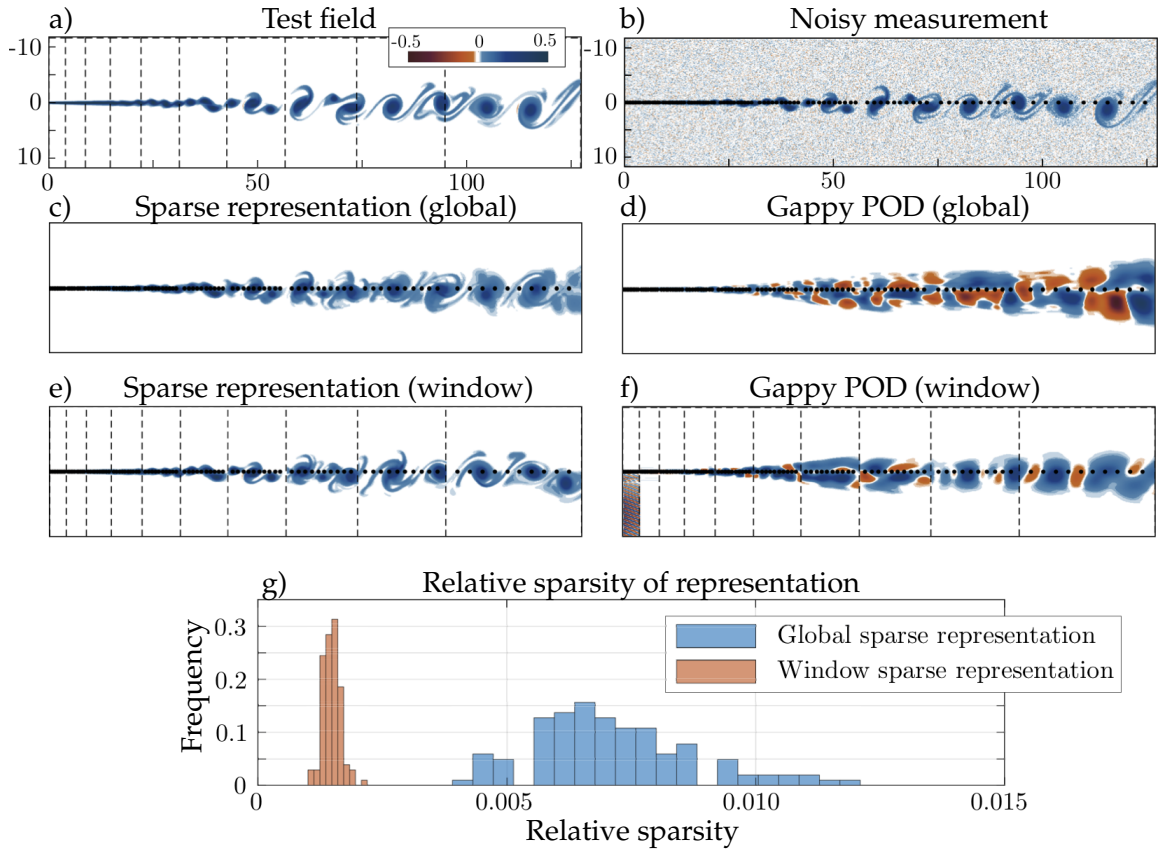

Figure~\ref{fig:mixing-downsample}g compares the relative sparsity of the global and windowed sparse representations, given by the fraction of total nonzero coefficients across all independent windowed optimization problems.  
 The local representations are more sparse and have higher fidelity than the global reconstruction.  
 This suggests a connection between multiscale features of the flow and the sparsity of representation.  
 By restricting the scope of the reconstruction problem, we simplify the effective dynamics, and this is reflected in the order of magnitude difference in the sparsity of representation.   
 These results suggest that the proposed flow field reconstruction method may generalize well to spatially complex flow fields.
 Note that on average, the various reconstruction results are comparable in an $\ell_2$ error metric, even though the fields obtained via sparse representations exhibit more visually accurate flow structures.  
 However, the $\ell_2$ metric is likely not ideal for measuring differences in convecting flow structures, as shifting the exact test field by  a couple of pixels will result in an error that is comparable to the gappy POD field.  
 

Figure~\ref{fig:mixing-centerline} demonstrates reconstruction from ten noisy point measurements evenly spaced along the centerline of each window.  In this case, sparse representation significantly outperforms gappy POD, although when averaged across the test data, the global sparse reconstruction is more accurate, in an $ \ell_2 $ sense, than the windowed estimate.  As discussed further in section~\ref{sec:discussion} and shown in figure~\ref{fig:pod-residual}, the dynamics have longer time scales in the downstream windows. 
Thus, the training data may not include enough representative examples of downstream behavior to admit a sparse representation. 
Indeed, the local reconstructions in the first seven windows are highly accurate and the errors in the total flow field estimate are largely due to the final three windows.

\subsection{Sea surface temperature field}\label{sec:results-sst}
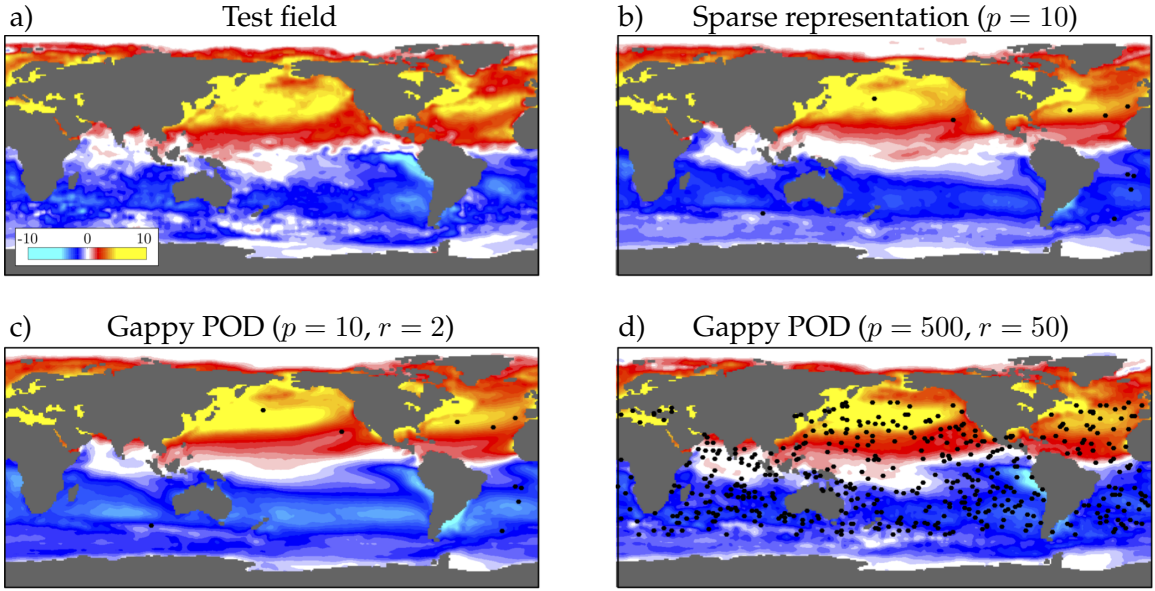
\begin{figure}
	\centering
	\begin{overpic}[width=.925\linewidth]{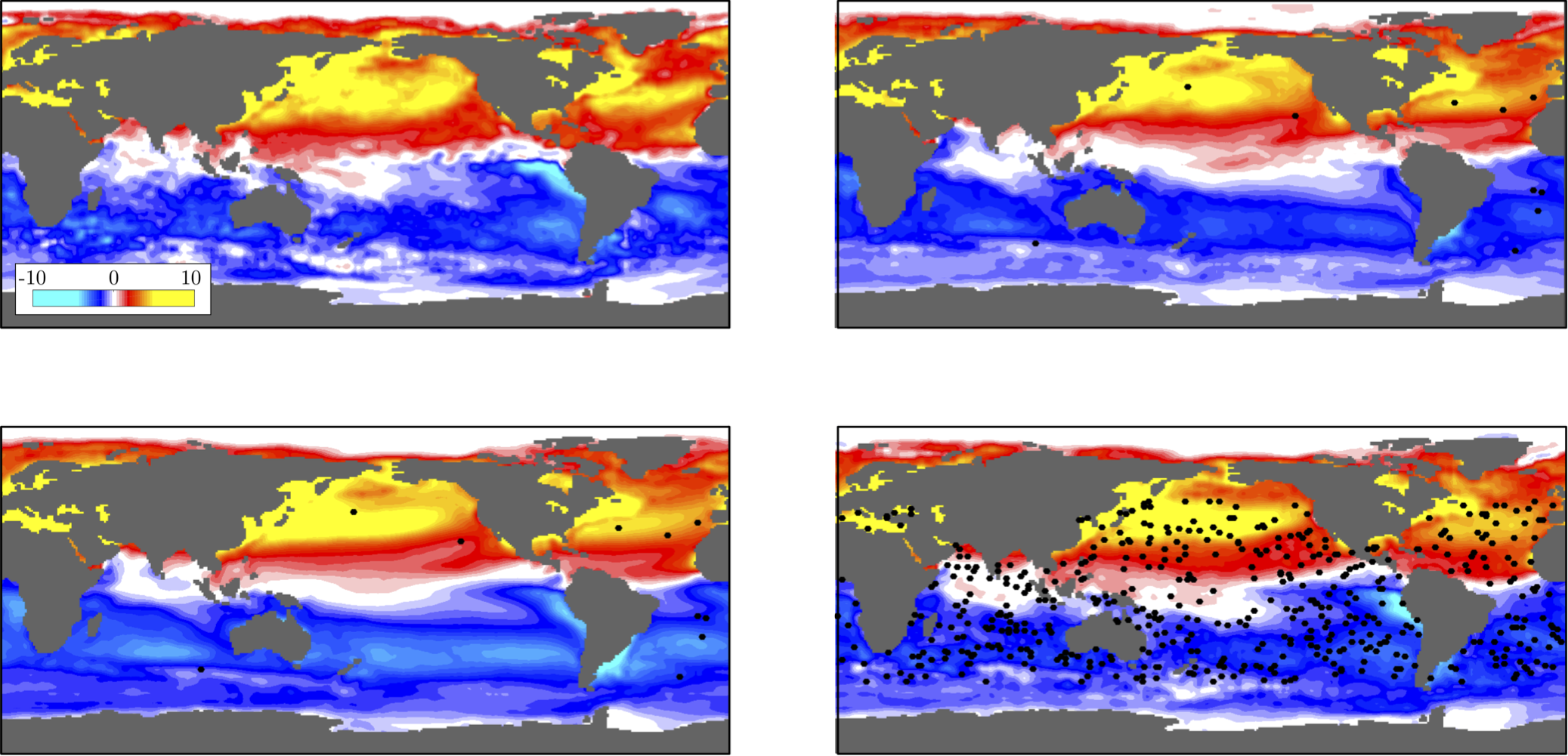}
		\put(19, 49){Test field}
		\put(60, 49){Sparse representation ($ p = 10 $)}
		\put(9, 22){Gappy POD ($ p = 10$, $r=2$)}
		\put(60, 22){Gappy POD ($ p = 500$, $r=50$)}
		\put(0.5, 49){a)}
		\put(53.5, 49){b)}
		\put(0.5, 22){c)}
		\put(53.5, 22){d)}
	\end{overpic}	
	\vspace{-.1in}
	\caption{Example reconstruction of a sea surface temperature field (a) using sparse representation in a training library (b) and gappy POD with $p=10$ and $p=500$ measurements (resp. c and d).  The POD libraries were truncated $r=2$ and $r=50$ modes respectively, which were the empirically determined optimal values.  Errors in all three estimates are around 0.30.  
		The long term annual mean temperature field has been subtracted to highlight variations in the data.}
	\label{fig:sst-reconstructions}
\end{figure}
The global sea surface temperature (SST) data set represents a flow field that is strongly driven by periodic seasonal forcing but also deviates from oscillatory behavior under the influence of complex oceanographic and environmental processes.  
For this reason, the SST data may be viewed as a problem of intermediate difficulty for the algorithm, with complexity somewhere between the $ \Re=100 $ flow past a cylinder from section~\ref{sec:results-cylinder} and the strongly aperiodic Gulf of Mexico data in section~\ref{sec:results-hycom}.  Figure \ref{fig:sst-reconstructions} shows a comparison of mean-subtracted temperature field reconstructions from randomly located point measurements restricted to the mid-latitude region from $ 50 ^\circ \text{S}$ to $50^\circ \text{N} $. We reconstruct the field with sparse representation in a training library and compare the result to gappy POD with the same measurements in a heavily truncated library ($r=2$) and with many more measurements in a library of $r=50$ POD modes. Both methods perform similarly in the error metric given by Eq. \eqref {eq:error-metric}; average reconstruction errors across the test data are within standard error of one another.

Fluctuations in the sea surface temperature field are dominated by seasonal oscillations, so that two POD modes capture a surprising amount of structure. In the absence of measurement noise, gappy POD has proven effective in estimating flow fields that can be represented accurately in terms only a few modes~\citep{Wilcox2006}. 
The fact that a flow as apparently complex as in figure~\ref{fig:sst-reconstructions}a can be accurately reconstructed with either method from as few as 10 random point measurements is a reflection of the underlying low-dimensional structure of this data set.  


\subsection{Gulf of Mexico surface vorticity}\label{sec:results-hycom}
\begin{figure}
	\centering
	\begin{overpic}[width=0.915\linewidth]{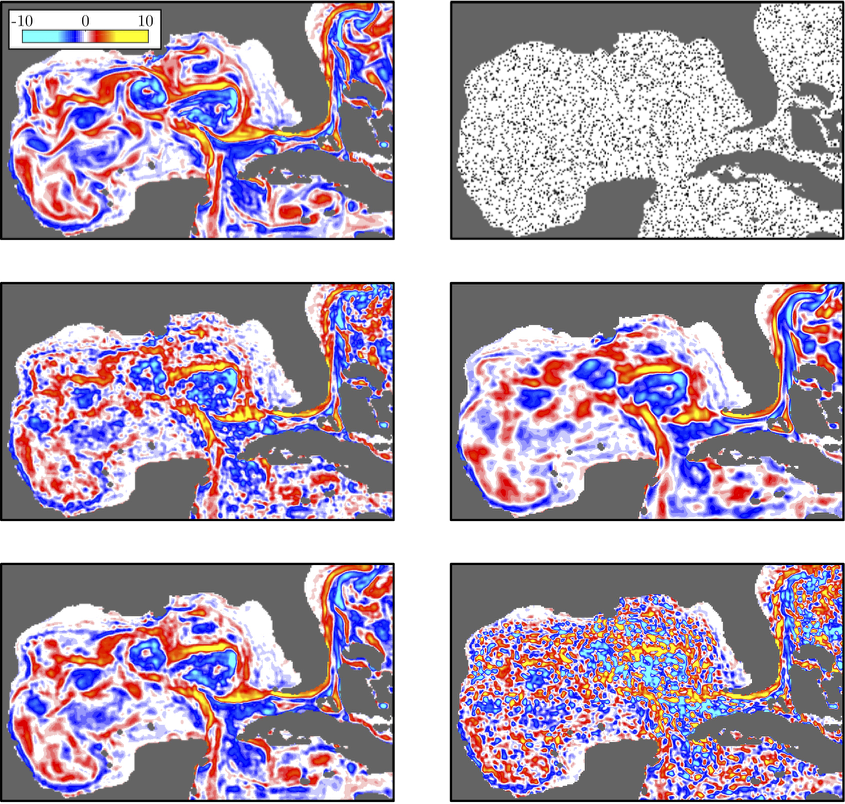}
		\put(0, 96){a)}
		\put(17, 96){Test field}
		\put(53, 96){b)}
		\put(63,96){Measurement locations}
		\put(0, 62.6){c)}
		\put(9, 62.6){Global sparse representation}
		\put(53, 62.6){d)}
		\put(65, 62.6){Global gappy POD}
		\put(0, 29.2){e)}
		\put(10, 29.2){Local sparse representation}
		\put(53, 29.2){f)}
		\put(66, 29.2){Local gappy POD}
	\end{overpic}
	\caption{Reconstruction of Gulf of Mexico vorticity field (a) from $ p = 4000 $ random points (b). The global sparse reconstruction from a training library (c) suffers from overfitting, since there is not a highly sparse global representation of this flow field in the training set.  Gappy POD on the global field (d) is comparable to global sparse representation; residual errors in both cases are $ \sim 0.60. $
	Sparse reconstruction from the same measurements, but using the local kernel method described in section \ref{sec:method-fluids} with $ k=96 $ equally spaced kernels, enables \textit{locally} sparse representations that combine to form a significantly more accurate global estimate (e), with a reconstruction error of 0.37.  The local least-squares POD still suffers from high-frequency overfitting (f). }
	\label{fig:hycom-reconstruction}
\end{figure}
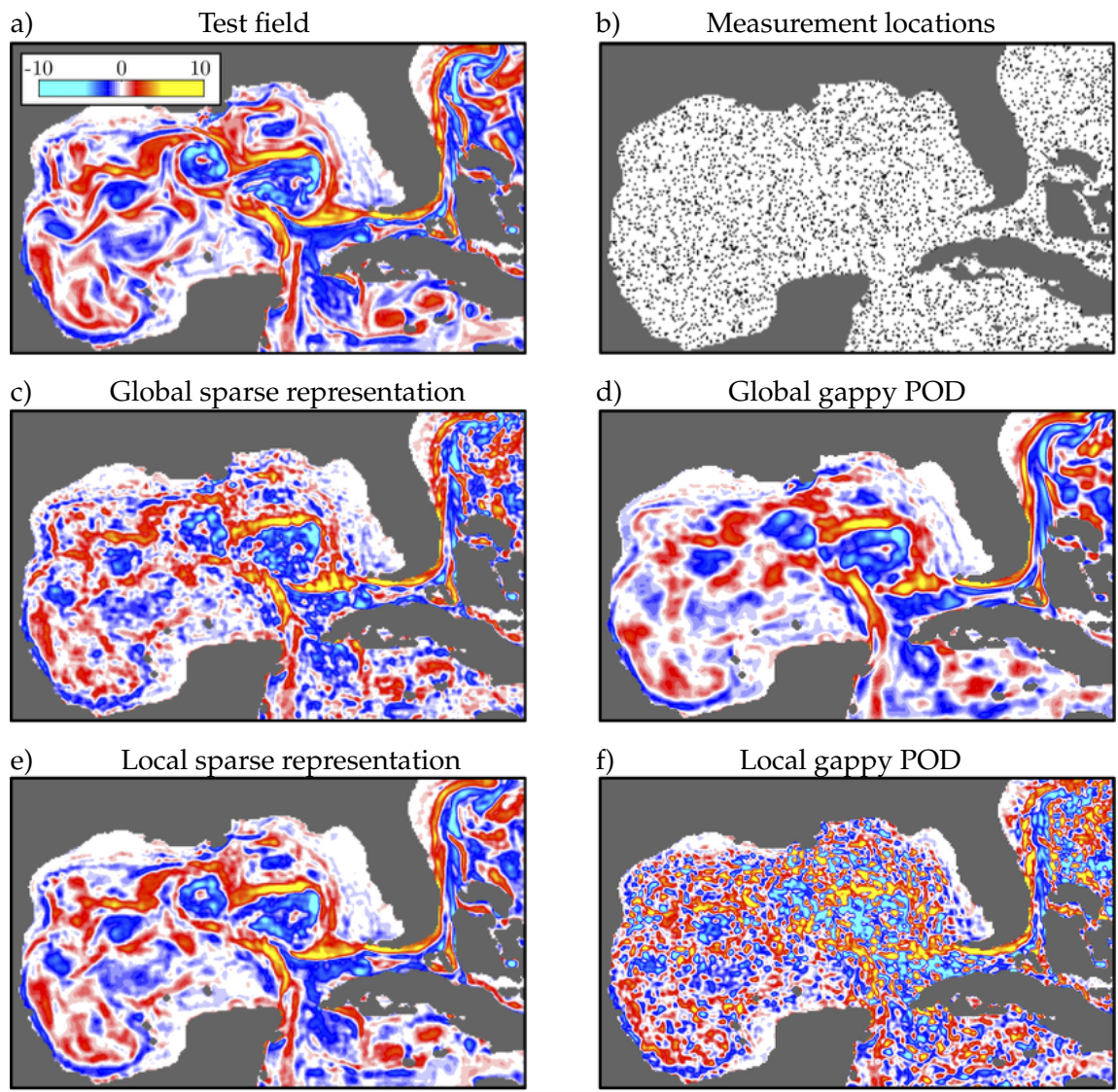

The final test flow is the HYCOM Gulf of Mexico ocean velocity data, which poses the greatest challenge for reconstruction.  On the time scale of the available data, the flow is not statistically stationary, and given the spatial complexity of the flow, accurate reconstruction requires significantly more measurements than the other fields considered in this work.  
Figure \ref{fig:hycom-reconstruction} shows a typical example of field reconstruction from $ p=4000 $ random point measurements, which account for about $ 8.5\% $ of all grid locations.  Global reconstructions from both sparse representation in the training set and gappy POD with a truncated library of $r=500$ modes successfully reconstruct much of the large-scale structure in the field and have comparable reconstruction errors, although the sparse representation estimate is contaminated by non-physical high-frequency fluctuations.

\begin{figure}
	\centering
	\begin{overpic}[width=0.915\linewidth]{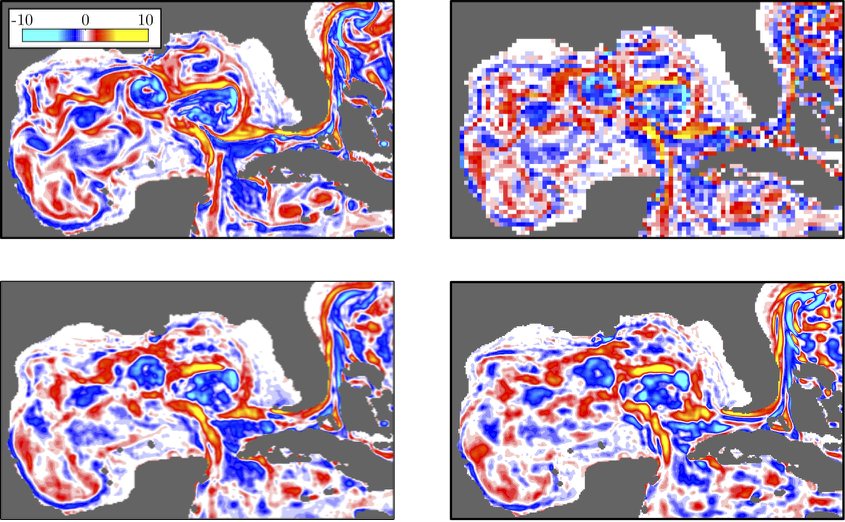}
		\put(0, 62.6){a)}
		\put(17, 62.6){Test field}
		\put(53, 62.6){b)}
		\put(67, 62.6){5:1 downsampling}
		\put(0, 29.2){c)}
		\put(9.5, 29.2){Local sparse representation}
		\put(53, 29.2){d)}
		\put(65, 29.2){Global gappy POD}
	\end{overpic}
	\vspace{-.1in}
	\caption{Reconstruction of Gulf of Mexico vorticity field (a) from uniform 5:1 downsampling (b). Local sparse representation (c) with $ k=96 $ equally spaced kernels yields a 26\% improvement in reconstruction error over gappy POD in a library truncated to $r=500$ modes (d). Global sparse representation and local gappy POD both perform worse than these methods. }
	\label{fig:hycom-downsample}
\end{figure}
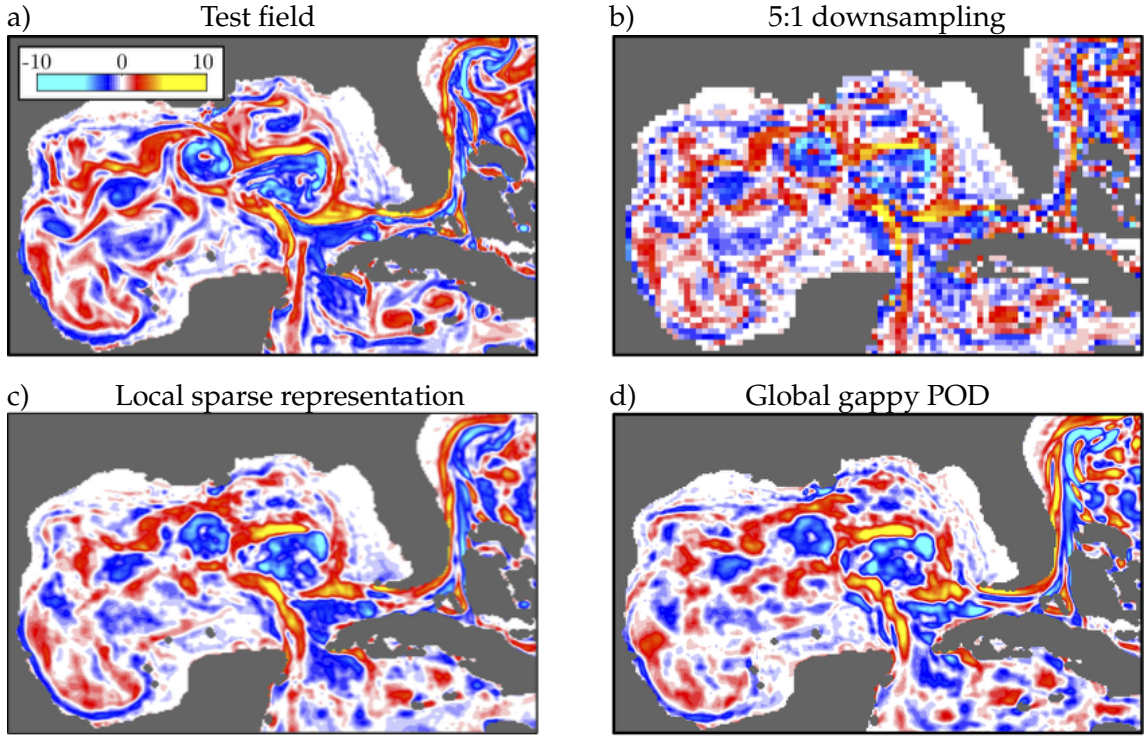

As with the mixing layer, the test field cannot be accurately represented as a sparse combination of training examples, because of the complexity of the flow and the fact that the training data does not fully generalize to the test set. The sparsest representation identified by equation \eqref{eq:sp-approx-noise}  contains $ K=3995 $ (around $ 48\% $) nonzero coefficients (figure \ref{fig:hycom-reconstruction}c), around 400 times as many as the sea surface temperature field in figure \ref{fig:sst-reconstructions}b. 
However, we achieve a more accurate reconstruction through the local kernel approach outlined in section \ref{sec:method-fluids} and the appendix.  By separating the reconstruction problem into localized kernels, we can stitch these local reconstructions together to obtain a global field that more accurately captures the large-scale vortical structures in the test field. 
This is intuitive from a measurement perspective, since the localization essentially relaxes the optimization constraints so that the sparse representation need only be consistent with \textit{local} measurements.  Seeking a global reconstruction that matches all measurements simultaneously is overly restrictive for a flow in which spatial correlations decay rapidly.  As with the mixing layer, sparsity appears to be a hallmark of complexity; on average, each kernel representation contains only $ K\approx 106 $ ($\sim 1\%  $) nonzero coefficients.  This suggests that features in local patches of the flow may closely resemble those present in the training data, even if the global flow field does not.

Figure \ref{fig:hycom-downsample} demonstrates reconstruction from uniformly downsampled measurements. 
The test field is sampled at a 5:1 ratio in the ocean region ($p=1261$ points) and reconstructed with gappy POD and local sparse representation, using the same parameters as in figure \ref{fig:hycom-reconstruction}.  Again, reconstruction from a local sparse representation is more accurate than gappy POD, suggesting that the method may be useful for interpolating low-resolution sensor data.

\section{Discussion}\label{sec:discussion}
This work demonstrates the enhanced robustness and accuracy of flow field reconstruction by sparse representation in a library of training data, and we have explored this approach on a range of example flow fields of increasing complexity.  
We also discuss potential limitations of sparse representation, along with proposed methodological extensions and improvements.  
The success of sparse representation depends on the availability of both an extensive library that contains representative examples of relevant flow structures and sufficiently rich sensor information to infer which of these structures are active.  
Both of these requirements are related to the flow physics and spatiotemporal scales of the particular system under consideration. 

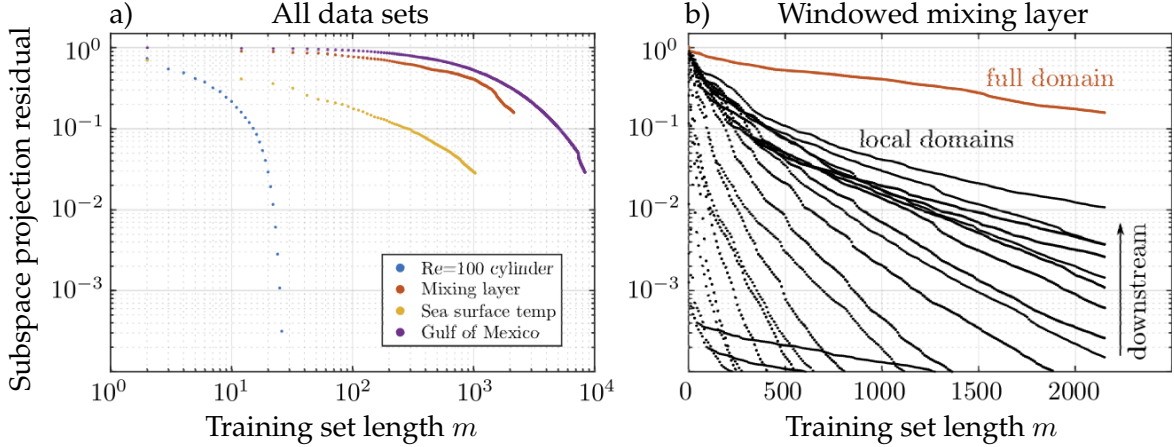
\begin{figure}
	\centering
	\vspace{0.2in}
	\begin{overpic}[width=0.9\linewidth]{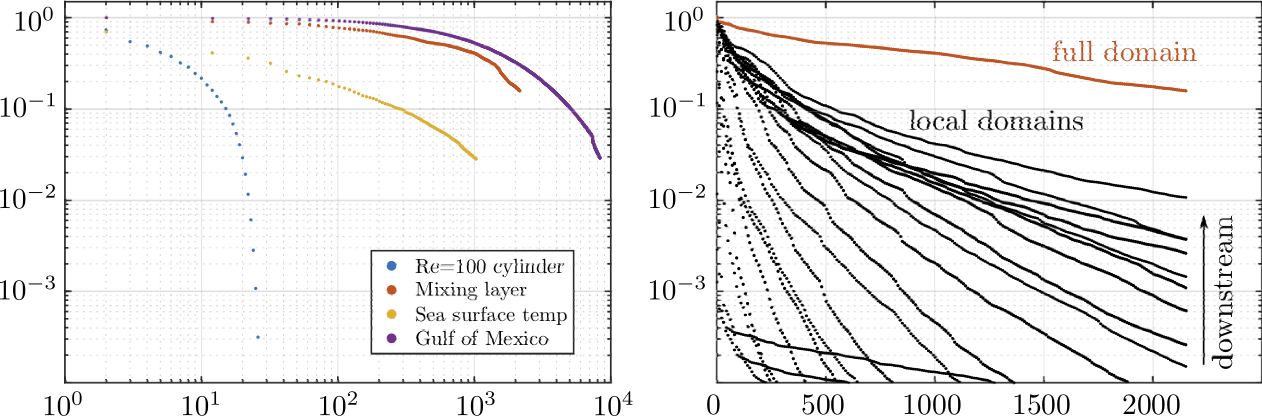}
		\put(5, 34){a)}
		\put(19.5, 34){All data sets}
		\put(56.5, 34){b)}
		\put(64.8, 34){Windowed mixing layer}
		\put(13.6,-3){Training set length $m$}
		\put(65.5,-3){Training set length $m$}
		\put(-4,1){\begin{sideways}Subspace projection residual\end{sideways}}
	\end{overpic}
	\vspace{0.15in}
	\caption{Residual error in projection of test data onto the linear subspace spanned by POD modes. As more data is added to the training set (horizontal axis), arbitrary fields from the test set are more likely to be in the span of the training data. This residual represents a lower bound on the error in a global reconstruction based on this linear subspace. (a) Comparison of all flows studied in this work. 
	(b) Subspace projection residuals for the windowing scheme shown in figures \ref{fig:mixing-downsample} and \ref{fig:mixing-centerline}. Even for windows encompassing the complex vortex pairing behavior downstream, the test data is approximately within the span of the subspace, although the time scales of convergence are much longer.}
	\label{fig:pod-residual}
\end{figure}

First, the library must contain a sufficiently extensive collection of example flow fields, so that a new flow field may be approximated by a sparse combination of these examples.  
Even for aperiodic flows, this may be satisfied if the training set contains a long enough flow history.  
To quantify if the training library is sufficiently complete to generalize to a new test field, we compute the residual error obtained by approximating a test field by orthogonal projection onto the training library.  
If the test field is well approximated in the training library, the residual is small, and if the test field has new structures that are not observed in the training data, there will be a large residual.  
Figure~\ref{fig:pod-residual} shows the residual error for each of the four flow examples from section~\ref{sec:results} as a function of the length of the training data; the orthogonal projection is obtained by computing the POD subspace for the given library with $m$ training examples.  
Even with very few examples in the library, the flow past a cylinder generalizes to the test data, since the flow is periodic.  
However, the mixing layer and Gulf of Mexico vorticity data have relatively large generalization error, even for large training libraries, indicating that there are new structures that haven't been observed in the training data.  
With enough training data, it is conceivable that the generalization error can be controlled for these flows, although this may be prohibitively expensive in terms of data collection and processing.  
Instead, decomposing the flow domain into local patches results in considerably improved library generalization (see figure \ref{fig:pod-residual}b), meaning that less training data is required for an accurate representation of a new flow field in the library.  
As shown in figures~\ref{fig:mixing-downsample},~\ref{fig:mixing-centerline} and~\ref{fig:hycom-reconstruction}, the local patches also admit a \emph{sparser} representation in the library, resulting in more accurate and robust flow reconstructions.  
The improved performance of a local sparse representation is intuitive, as decomposing the spatial domain makes it more likely to find similar local flow structures in the training data.  
Thus, the generalization error of the training library provides a useful diagnostic to quantify the expected performance of sparse representation.  

Throughout these examples, we find that least-squares solutions generally overfit to noisy sensor measurements, resulting in non-physical high-frequency fluctuations, as in figures \ref{fig:mixing-downsample} and \ref{fig:hycom-reconstruction}. 
In contrast, sparse representation in a library of training examples results in robust and accurate flow reconstruction, preventing overfitting and ensuring that that the unmeasured regions of the field are consistent with prior knowledge. 
If a sufficiently sparse approximation is not possible, however, the method will not reliably produce a field that is qualitatively similar to actual observations. 
However, even if a sparse representation of the entire flow field does not exist, figures \ref{fig:mixing-centerline}b-c and \ref{fig:hycom-reconstruction}d demonstrate that locally sparse representations can be used for globally accurate reconstruction.

\begin{figure}
	\centering
	\begin{overpic}[width=0.615\linewidth]{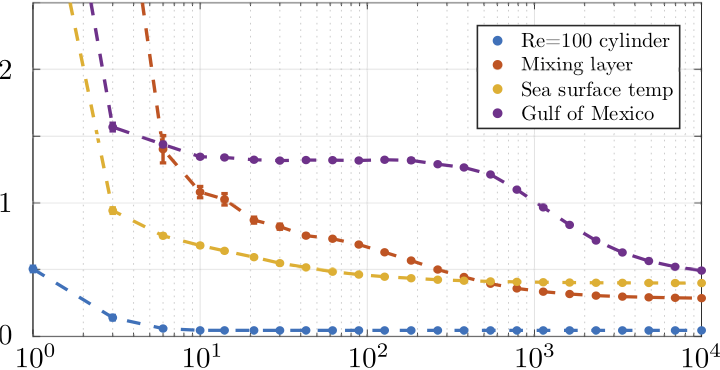}
		\put(30,-4){\# random measurements $n_s$}
		\put(-5,12){\begin{sideways}Reconstruction error\end{sideways}}
	\end{overpic}
	\vspace{0.15in}
	\caption{Normalized residual error of sparse representation-based reconstructions with increasing number of random point measurements. We use the global reconstruction method for the periodic vortex shedding past a cylinder (blue), mixing layer (red), sea surface temperature fields (yellow), and Gulf of Mexico (purple) flow fields. For computational efficiency we estimate coefficient vectors $ \boldsymbol{\hat{s}} $ using Orthogonal Matching Pursuit (OMP) with empirical estimates of the sparsity $ K. $ Accurate reconstruction requires both rich training data and sufficient measurement information; the latter condition can vary widely depending on the flow.}
	\label{fig:err-vs-ns}
	\vspace{-.1in}
\end{figure}

A second condition for successful sparse flow reconstruction, even with a rich enough library, is for the measurements to provide sufficient information to correctly identify the sparse library coefficients.  
For example, it is unrealistic to hope that a single point measurement can be used to reconstruct a highly turbulent flow field, no matter how comprehensive the training library. In the Bayesian perspective, even perfect knowledge of the prior distribution is not enough for an accurate estimation, unless it is sufficiently conditioned on measurement information.
Figure \ref{fig:err-vs-ns} shows the accuracy of the sparse representation method versus the number of random point measurements for each example.  
In each case, there is a rough number of sensors where the error sharply decreases: $ p = 10 $ for the flow past a cylinder and sea surface temperature fields, $ p = 100 $ for the mixing layer, and $ p = 4000 $ for the Gulf of Mexico data, which roughly correspond to the number of measurements used in section~\ref{sec:results}.  
Further increasing the number of sensors results in a minimum error plateau, which is defined by the generalization error of the library, as described above. 
It is important to note that it may be possible to reconstruct the flow field with less error from fewer measurements by leveraging additional knowledge about the flow, for example from a reduced-order model \citep{Rowley2017} or via time-delay embedding \citep{Brunton2017}. 

\section{Conclusion}\label{sec:conclusion}
In this study, we develop a method for flow field reconstruction based on sparse representation in a library of examples.  This method builds on prior work in library-based reconstruction and sparse representation, in particular the sparse representation for classification algorithm~\citep{Wright2009,Bright2013}.  
We apply this method to several example flows, ranging from simple canonical flows to challenging geophysical data sets, and demonstrate improved accuracy and robustness to noise and corruption compared to typical least-squares flow reconstruction, provided that the library is sufficiently rich and the measurements are sufficiently informative.  

This work suggests several directions of ongoing research to refine the method for practical applications.  
Although figures \ref{fig:pod-residual} and \ref{fig:err-vs-ns} give rough metrics for sufficiency of the training data and measurement information, more quantitative and principled criteria for both requirements would be useful to determine \textit{a priori} which flows are good candidates for this method, how much training data is required, and which sensor configurations will sufficiently inform the structure of the flow field.  
The performance of this method may also be improved with more sophisticated library learning methods \citep{Aharon2006, Elad2006, Mathelin2018} or optimal sensor placement strategies \citep{Wilcox2006, Yildrim2009, Manohar2017, Mons2017}, both of which are active areas of research.  
Further, there are many ways to formulate and solve the sparse optimization problem \eqref{eq:sp-approx-noise}, and alternative approaches, such as sequentially thresholded least-squares~\citep{Zheng2018arxiv} or pursuit algorithms, may outperform $ \ell_1 $ regression. 
In addition, the sparse optimization procedure may be too computationally expensive for some real-time control applications, motivating ongoing work to improve algorithmic efficiency; fortunately, the timescales of the geophysical flows investigated in this work are slow compared with the sparse optimization. 
Finally, most flows of interest are three-dimensional, and it will be important to demonstrate this method on three-dimensional flows.  
While it is straightforward to generalize this method to three-dimensional fields, e.g., by the same vectorization approach used for two-dimensional flows, the matrices representing discretized 3D flows can become very large.  
The main computational cost is due to the optimization problem, which scales with the number of measurements and not the dimensionality of the discretized flow field; however, the number of measurements necessary to sufficiently inform the structure of a complex three-dimensional flow field may lead to prohibitively expensive computations.


In the examples considered here, we have shown several advantages which make sparse representation an attractive candidate for flow field reconstruction. In particular we find that reconstruction from sparse representation in a library of training examples is robust to noise and leads to physical flow field estimates. We have shown that this framework can be modified to handle dense sensor noise and gross measurement corruption.  The method can also be extended to complex flow fields by decomposing the spatial domain and seeking localized sparse representations. With this flexibility, sparse representation may provide a powerful tool for estimating complex flow fields in a range of applications.

\section*{Acknowledgements} 
We gratefully acknowledge funding support from the Air Force Office of Scientific Research (FA9550-18-1-0200 \& FA9550-16-1-0650) and the Army Research Office (W911NF-17-1-0422).  KM gratefully acknowledges support by the Washington Research Foundation, the Gordon and Betty Moore Foundation (Award \#2013-10-29), the Alfred P. Sloan Foundation (Award \#3835), and the University of Washington eScience Institute.  
Special thanks to Nathan Kutz for sharing his insights into the extreme utility of sparsity promoting methods for complex physical systems.  We also thank Bing Brunton, Ben Erichson, and Lionel Mathelin for valuable discussions on sparsity and flow estimation. 

\section*{Appendix: local reconstruction method}

Here, we provide details on the kernel-based localized reconstruction method applied to the mixing layer and Gulf of Mexico vorticity fields, for example to produce the estimates in figure \ref{fig:hycom-reconstruction}e-f.  
As described in section \ref{sec:method-fluids}, we construct a global flow field estimate as a weighted superposition of local reconstructions in a decomposed domain, which admits sparser representations in the training library.

We introduce compact overlapping kernels $ \boldsymbol{\Phi}_j, \; j=1, 2, \dots, k $ normalized so that at each grid location $ \mathbf{r}, $ $ \sum_j \boldsymbol{ \Phi}_j(\mathbf{r}) = 1. $  These kernels separate the global estimation problem into $ k $ local problems:\begin{equation*}
\boldsymbol{\hat{s}}_j = \arg \min_{\boldsymbol{s}_j} \| \boldsymbol{s}_j\| _q \hspace{1cm} \text{subject to} \hspace{1cm} \| \boldsymbol{y}_j - \boldsymbol{C}\boldsymbol{ \Phi}_j \boldsymbol{\Psi s}_j\| _2 < \epsilon.
\end{equation*} Since the kernels are normalized, the local flow field estimates $ \boldsymbol{\hat{x}}_j = \boldsymbol{\Phi}_j\boldsymbol{\Psi \hat{s}}_j $ can be combined to form a global estimate $ \boldsymbol{\hat{x}} = \sum_j \boldsymbol{\hat{x}}_j. $  The simplest such kernels are the windows used for the mixing layer reconstructions in figure \ref{fig:mixing-centerline}c-d, where each kernel has the value 1 within its window and 0 outside.

\begin{figure}
	\centering
	\vspace{.15in}
	\begin{overpic}[width=0.6\linewidth]{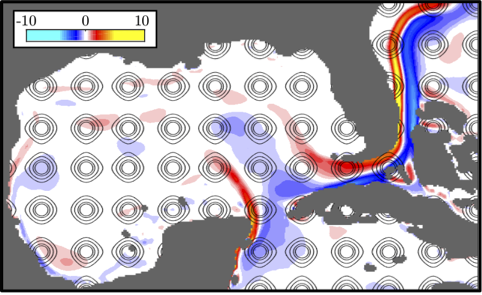}
	\end{overpic}
	\caption{Locations of kernel centers for the local reconstructions in figure \ref{fig:hycom-reconstruction}e-f, shown superimposed on the empirical mean vorticity field.  The contours shown are for Gaussian kernels prior to normalization.}
	\label{fig:hycom-kernels}
\end{figure}

For the Gulf of Mexico data we define 96 points $ \mathbf{r}_1, \; \mathbf{r}_2, \dots, \; \mathbf{r}_j, \; \mathbf{r}_{96}$ as the kernel centers on a uniform $12 \times 8$ grid covering the spatial domain, as in figure \ref{fig:hycom-kernels}.  Each kernel $ \boldsymbol{ \Phi}_j $ is constructed with a radial Gaussian function of the distance from each grid location to the kernel center: \begin{equation*}
\boldsymbol{\Phi}_j(\mathbf{r}) = \frac{1}{N(\mathbf{r})}e^{|\mathbf{r} - \mathbf{r_j}|^2/\sigma^2},
\end{equation*} where the width $ \sigma $ is half the longitudinal distance between successive kernel centers. Values below $ 10^{-2} $ are set to zero, and the normalization factors $ N(\mathbf{r}) $ are then calculated as the sum of the unweighted values of all kernels \begin{equation*}
N(\mathbf{r}) = \sum_{j=1}^{k} \boldsymbol{ \Phi}_j(\mathbf{r}),
\end{equation*} so that the resulting estimates may be combined in a weighted average.

Computationally, each location $ \mathbf{r} $ is a point on the grid, and so just as the flow field snapshots are arranged into column vectors, the corresponding values of $ \boldsymbol{ \Phi}_j $ are the entries in a sparse diagonal matrix. The product $ \boldsymbol{\Phi}_j\mathbf{x} $ of a kernel with a discretized flow field is then nonzero only in the region surrounding the kernel center $ \mathbf{r}_j. $

\section*{Appendix B: Parameter tuning}

Machine learning methods typically allow for tuning some set of parameters for optimal performance.  For example, increasing the regularization parameter $ \lambda $ in Eq.~\eqref{eq:overdetemined-regularized} can prevent overfitting to noisy data, but beyond some optimal value the solution is prone to bias.  For the sparse representation problem~\eqref{eq:sp-approx-noise} which is central to our proposed method, the only free parameter is the relaxation $ \epsilon$.  A larger value of $ \epsilon $ allows for a larger difference between the observations and the estimated flow field, which may be useful for instance if the measurements are noisy or if the training data may not generalize well.  In these cases, relaxing the constraint can lead to a sparse solution and accurate estimation.

\begin{figure}
	\centering
	\vspace{.15in}
	\begin{overpic}[width=0.8\linewidth]{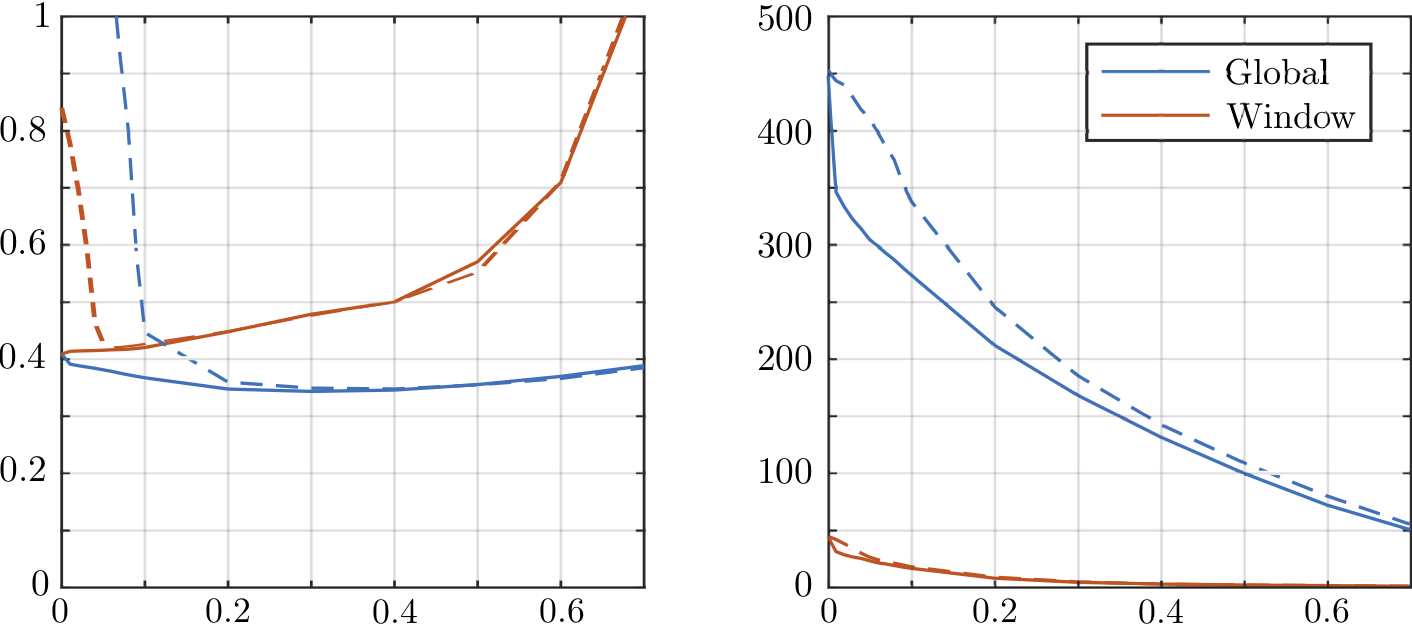}
		\put(5, 44){(a)}
		\put(10, -3){Relaxation parameter $ \epsilon $}
		\put(-3.5, 9.5){\begin{sideways}Reconstruction error\end{sideways}}
		\put(58.5, 44){(b)}
		\put(65, -3){Relaxation parameter $ \epsilon $}
		\put(50, 9.5){\begin{sideways}Average sparsity $ K $\end{sideways}}
	\end{overpic}
	\vspace{.15in}
	\caption{The sparse representation problem in Eq.~\eqref{eq:sp-approx-noise} allows for choosing the relaxation parameter $ \epsilon $. (a) Average reconstruction error across the test set vs. $ \epsilon $ for the downsampled mixing layer (see figure \ref{fig:mixing-downsample}) from clean and noisy measurements (solid and dashed lines, respectively). (b) Sparsity of representation for the same reconstruction problem vs. $ \epsilon $. Relaxing the problem improves reconstruction accuracy, especially when the measurements are noisy, but increasing $ \epsilon $ too far can allow for solutions which are inconsistent with measurements.  }
	\label{fig:mixing-relaxation}
\end{figure}

The sparse representation method can be sensitive to the choice of $ \epsilon. $ For example, figure~\ref{fig:mixing-relaxation}a shows the error in reconstructing a downsampled test field (see figure~\ref{fig:mixing-downsample}) with increasing relaxation. For clean data and global reconstruction (solid blue lines), the method is only weakly dependent on the choice of $ \epsilon $, whereas a careful selection of this parameter is important for reconstruction from windows (red lines) and if measurements are noisy (dashed lines).  For the windowed reconstruction, error increases sharply beyond an optimal value of $ \epsilon $; figure~\ref{fig:mixing-downsample}b suggests that this may be because overly relaxed constraints allow the solution to be highly sparse, but generally inconsistent with observations.  We expect that an appropriate choice of $ \epsilon $ will in general also depend on both the magnitude of observed fluctuations and the number of measurements.

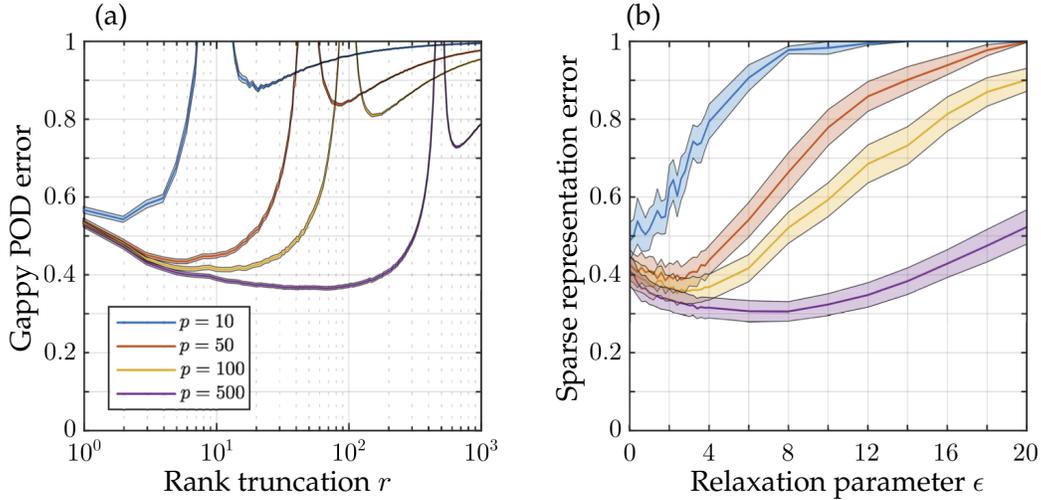
\begin{figure}
	\centering
	\vspace{.15in}
	\begin{overpic}[width=0.8\linewidth]{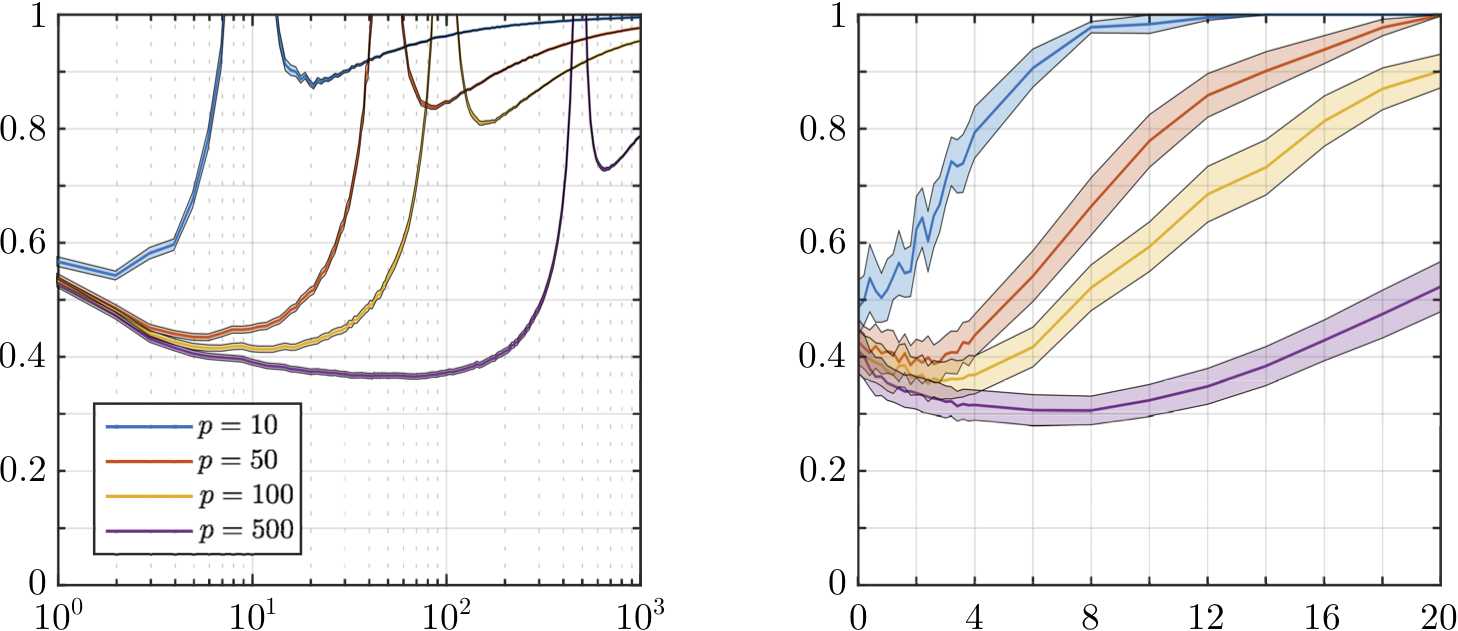}
		\put(5, 44){(a)}
		\put(12, -3){Rank truncation $ r $}
		\put(-3.5, 11){\begin{sideways}Gappy POD error\end{sideways}}
		\put(58.5, 44){(b)}
		\put(65, -3){Relaxation parameter $ \epsilon $}
		\put(51.5, 6){\begin{sideways}Sparse representation error\end{sideways}}
	\end{overpic}
	\vspace{.15in}
	\caption{Parameter tuning for sea surface temperature field estimation. (a) Rank truncation of gappy POD library.  Sharp peaks correspond to $ p = r, $ when the number of point measurements equals the number of retained modes  and the matrix $ \boldsymbol{C\Psi} $ becomes square.  The optimal truncation appears to depend on the number of measurements, but is always in the oversampled case $ r < p. $ (b) Relaxation of the sparse representation problem.  As with the mixing layer, some relaxation of the constraint can improve accuracy, but an overly relaxed optimization problem leads to inconsistent estimates.  For both plots, shaded bands indicate standard error on the mean.   }
	\label{fig:sst-truncation}
\end{figure}

Gappy POD can also be tuned via truncation of the POD library.  In this case the optimization problem is typically the overdetermined case~\eqref{eq:overdetermined-L2}, although if there are fewer measurements than modes in the library, the least-squares solution to~\eqref{eq:underdetermined-approx} with $ q=2 $ is an analogous estimation method.  We find that gappy POD performs best in the oversampled case $ r < p $. That is, the library is truncated to contain fewer modes than measurements. For example, gappy POD has comparable accuracy to sparse representation for the sea surface temperature field estimates in figure \ref{fig:sst-reconstructions}, but this accuracy is sensitive to the truncation. Figure~\ref{fig:sst-truncation}a shows gappy POD reconstruction error vs. rank truncation $ r $ averaged across the test data for various numbers of random point measurements $ p $.  The error peaks sharply when the measured library matrices $ \boldsymbol{C\Psi} $ become nearly square,  
but for any value of $ p $ the optimal truncation is $ r<p $.  

Figure~\ref{fig:sst-truncation}b shows reconstruction error vs. relaxation parameter $ \epsilon $ for sparse representation-based estimates of the sea surface temperature fields for varying number of point measurements.  The points are chosen at random with a new realization for each field in the test set.  In this case we do not artificially introduce noise, although an appropriate choice of $ \epsilon $ still improves the accuracy of the sparse representation method.

\setlength{\bibsep}{3.5pt plus 1ex}
\begin{spacing}{.01}
\small
\bibliographystyle{unsrt}
\bibliography{refs}
\end{spacing}

\end{document}